\documentclass[10pt,twocolumn,twoside]{IEEEtran}
\IEEEoverridecommandlockouts 
\usepackage{amsthm,amsmath,amssymb}
\usepackage{arydshln}
\usepackage{color}
\usepackage{tikz,pgf,pgfplots}
\usepackage{tkz-graph}
\usetikzlibrary{arrows}
\usetikzlibrary{arrows.meta}
\usepackage{subfigure}
\tikzset{every picture/.style={scale=0.52}} 
\pgfplotsset{compat=1.18}

\usepackage{multirow} 

\definecolor{mycolor1}{HTML}{DCE125}%
\definecolor{mycolor2}{HTML}{5086C4}%
\definecolor{mycolor3}{HTML}{B55489}%
\definecolor{mycolor4}{HTML}{4C6C43}%
\definecolor{mycolor5}{HTML}{F0A19A}%
\definecolor{mycolor6}{HTML}{7C7CBA}%
\definecolor{mycolor7}{HTML}{00A664}%
\definecolor{mycolor8}{HTML}{F9ED1D}%
\definecolor{mycolor9}{HTML}{3FA0C0}%

\usepackage{algorithm}
\usepackage{algpseudocode}

\newcommand{\C}{\mathbb{C}}
\newcommand{\R}{\mathbb{R}}
\newcommand{\N}{\mathbb{N}}

\newcommand{\cG}{\mathcal{G}}

\newcommand{\cN}{\mathcal{N}}
\newcommand{\cO}{\mathcal{O}}
\newcommand{\cP}{\mathcal{P}}
\newcommand{\cQ}{\mathcal{Q}}

\newcommand{\cU}{\mathcal{U}}
\newcommand{\cV}{\mathcal{V}}
\newcommand{\cW}{\mathcal{W}}

\newcommand{\ave}{\textnormal{ave}}
\DeclareMathOperator*{\diag}{diag}
\DeclareMathOperator*{\ran}{ran}

\DeclareMathOperator{\real}{Re}
\DeclareMathOperator{\imag}{Im}

\DeclareMathOperator*{\esssup}{ess\,sup}
\newcommand{\s}{\textnormal{s}}

\newcommand{\Lc}{\hat L_{\textnormal{c}}}
\newcommand{\Nc}{\cN_{\textnormal{c}}}

\newcommand{\Rp}{\R_{\geq 0}}

\DeclareMathOperator*{\cond}{cond}

\newtheorem{definition}{Definition}

\newtheorem{theorem}{Theorem}
\newtheorem{lemma}{Lemma}
\newtheorem{corollary}{Corollary}
\newtheorem{proposition}{Proposition}
\newtheorem{remark}{Remark}
\newtheorem{problem}{Problem}

\allowdisplaybreaks

\title{Scalable Distributed Least Squares Algorithm for Linear Algebraic Equations via Periodic Scheduling\thanks{This work was supported by the National Natural Science Foundation of China under grants 62203053 and 62261160575.}}

\author{Shenyu Liu${}^*$\thanks{S. Liu is with the School of Automation, Beijing Institute of Technology, China, \texttt{shenyuliu@bit.edu.cn}.}}

\begin{document}
	
	\maketitle
	
	\begin{abstract}
		In this work, we propose a novel discrete-time distributed algorithm for finding least-squares solutions of linear algebraic equations with a scheduling protocol to further enhance its scalability. Each agent in the network is assumed to know some rows of the coefficient matrix and the corresponding entries in the observation vector. Unlike typical distributed algorithms, our approach considers communication bandwidth limits, allowing agents to transmit only a portion of their ``guessed" solution, independent of its dimension. A {\color{blue}cyclic} scheduling protocol determines which portion is transmitted at each iteration. Assuming a small fixed step size and a diagonalizable algorithm matrix, we prove that agents' ``guessed" solutions converge exponentially to a least squares solution. For cases where the observation vectors are time-varying, a modified algorithm guarantees practical convergence, with tracking error bounded by the single-step variation in the observation vector. {\color{blue}Simulations and comparisons with state-of-the-art algorithms validate our algorithm's feasibility and scalability.}
	\end{abstract}

	\section{Introduction}\label{sec:intro}
	
	Solutions to linear algebraic equations (LAEs) find applications in diverse scientific and engineering disciplines such as physics, chemistry, financial modeling, and network analysis. A particular area of interest concerns data-driven modeling and regression, which typically employ very large-dimensional, data-dependent LAEs for prediction. Since the data can be distributed across multiple nodes in a network, an efficient, privacy-preserving, and dependable algorithm is crucial for training such models. To ensure this, the associated distributed algorithms must account for both the computational and communication limitations of the digital devices involved in the process~\cite{XC-TB-SD-YCE-KBL-HVP-JZ:23}.
	
	Applications of solutions to LAEs include parameter estimation in sensor networks~\cite{SK-JMFM-KR:12}, data fitting~\cite{PCH-VP-GS:13}, linear and nonlinear (kernel-based) regression in predictive learning~\cite{MA-PLB:99}, distributed spectrum estimation~\cite{SL-JC-SM:25}, and system identification~\cite{LL:99}. Early distributed algorithms for solving LAEs--such as the Jacobi, Gauss-Seidel, and Richardson methods--are well-documented in~\cite{DPB-JNT:97}. More recently, fully distributed algorithms have been developed, with works like \cite{SM-JL-ASM:15,BDOA-SM-ASM-UH:16,WL-XZ-YH-HJ:21} proposing solutions for cases where LAEs admit exact solutions, among others. 
	
	{\color{blue}The more challenging case of \emph{overdetermined} LAEs, where exact solutions do not exist, has driven the development of several distributed algorithms for least-squares (LS) solutions. By forming the problem of solving LAEs as an optimization problem where the objective function is a sum of convex functions, each known to an agent, the works \cite{JW-NE:10,BG-JC:14-tac} inspire continuous-time flows for finding LS solutions. The distributed algorithm proposed in \cite{GS-BDOA-UH:17s} ensures that the agents' states can converge to an arbitrarily small ball around a LS solution by selecting a sufficiently large gain for the ``consensus + projection" flow. To further achieve the LS solutions, the work~\cite{AN-AO-PAP:10tac,YL-YL-BDOA-GS:20} proposes diminishing step size algorithms, under which the discretization of the continuous-time algorithm asymptotically converges with a sublinear convergence rate of $\cO(1/t)$. Distributed LS algorithms with fixed step size and exponential convergence rate are studied in recent works~\cite{XW-JZ-SM-MJC:19,TY-JG-JQ-XY-JW:20,YL-CL-BDOA-GS:19}.} The problem of distributed solutions to LAEs is also extended to time-varying communication graphs~\cite{JL-ASM-AN-TB:17}, random communication graphs~\cite{SSA-NE:21}, and separable data~\cite{PS-JC:22-csl}. See also the survey~\cite{PW-SM-JL-WR:19} for a comprehensive overview of related works.
	
	The aforementioned distributed algorithms mainly assume that each agent in the network has access to some rows of the coefficient matrix of LAEs and the corresponding entries in the observation vector. Those algorithms then iterate the following two steps alternatively:
	\begin{enumerate}
		\item Each agent broadcasts a packet to each neighboring agent and receives packets from them, confined by a communication topology.
		\item Based on the received packets, its own known data, and the current state variables, each agent computes new state variables.
	\end{enumerate}
	Typically, the packets for transmission include the agents' estimates of LS solutions, and thus their sizes are proportional to the solution's dimension. {\color{blue}This poses a significant challenge due to the limited bandwidth for communication among agents, as the packets may require multiple clock cycles for transmission. As illustrated in Fig.~\ref{fig:comm}, agents must wait through multiple clock cycles per iteration until they receive the complete packets from their neighbors before proceeding with computation. This results in a high \emph{communication-to-computation} ratio, leaving agents idle for considerable periods during each iteration. From another perspective, there's wasted potential in each iteration that leads to algorithmic inefficiency, and the algorithm's efficiency decreases as the solution dimension increases.} The bandwidth limitation renders most of the aforementioned distributed algorithms not \emph{scalable}, as the resulting inefficiency cannot be mitigated by simply increasing the number of agents in the system. Consequently, an efficient algorithm should achieve a balance between clock cycles allocated to computation and communication. In other words, a scalable algorithm should transmit packets with sizes that are independent of the solution's dimension among the agents per iteration.
	
	\begin{figure}
		\centering
		\subfigure[Multiple clock cycles of communication per iteration]{\begin{tikzpicture}[scale=1.5]
	\draw[-{Stealth},line width=1.5pt] (0,0) -- (8,0);
	\draw[-{Stealth},line width=1.5pt] (0,2) -- (8,2);
	
	\foreach \x in {0,...,3}{			
		\foreach \xx in {1,...,4}{
			\coordinate (\x\xx1) at (1.7*\x-0.3*\xx+1.7,2);
			\coordinate (\x\xx2) at (1.7*\x-0.3*\xx+1.7,0);
			
			\draw[{Stealth}-,color=mycolor2,line width=1.5pt] (\x\xx1) -- +(-0.2,-2);
			\draw[{Stealth}-,color=mycolor3,line width=1.5pt] (\x\xx2) -- +(-0.2, 2);
		}
		\draw[fill=green,color=mycolor7] (1.7*\x+1.4,2) rectangle +(0.6,1);	
		\draw[fill=green,color=mycolor7] (1.7*\x+1.4,0) rectangle +(0.6,-1);	
		
		\draw[fill=yellow,color=mycolor8] (1.7*\x+1.4,2) rectangle +(-1.1,1);	
		\draw[fill=yellow,color=mycolor8] (1.7*\x+1.4,0) rectangle +(-1.1,-1);	
		\draw[dashed] (0.3+1.7*\x,-1) -- (0.3+1.7*\x,3);		
	}
	\node[] at (-.5,2.5) {Agent 1};
	\node[] at (-.5,-.5) {Agent 2};
	\node[] at (8, 1) {Time};
	
\end{tikzpicture}\label{subfig:slow}}
		\subfigure[Single clock cycle of communication per iteration]{\begin{tikzpicture}[scale=1.5]
	\draw[-{Stealth},line width=1.5pt] (0,0) -- (8,0);
	\draw[-{Stealth},line width=1.5pt] (0,2) -- (8,2);
	
	\foreach \x in {0,...,7}{
		\coordinate (\x1) at (0.8*\x+0.7,2);
		\coordinate (\x2) at (0.8*\x+0.7,0);
		
		\draw[{Stealth}-,color=mycolor2,line width=1.5pt] (\x1) -- +(-0.2,-2);
		\draw[{Stealth}-,color=mycolor3,line width=1.5pt] (\x2) -- +(-0.2, 2);
		
		\draw[fill=green,color=mycolor7] (\x1) rectangle +(0.6,1);	
		\draw[fill=green,color=mycolor7] (\x2) rectangle +(0.6,-1);	
		
		\draw[fill=yellow,color=mycolor8] (\x1) rectangle +(-0.2,1);	
		\draw[fill=yellow,color=mycolor8] (\x2) rectangle +(-0.2,-1);

		\draw[dashed] (0.5+0.8*\x,-1) -- (0.5+0.8*\x,3);		
	}
	\node[] at (-.5,2.5) {Agent 1};
	\node[] at (-.5,-.5) {Agent 2};
	\node[] at (8,1) {Time};
\end{tikzpicture}\label{subfig:fast}}
		\caption{Temporal sequences of communication and computation for two agents. Packet transmission between the two agents are depicted by magenta and blue arrows. {\color{blue}Because the agents cannot proceed with computation (green) before they receive the complete packets, they remain idle (yellow) during communication.} Vertical dashed lines indicate the beginning of each iteration. Over the same time interval, the scheme (a) processes 8 iterations of computation while the scheme (b) processes 4 iterations of computation.}\label{fig:comm}
	\end{figure}
	
	{\color{blue}One way to better balance computation and communication is through a double-layered network~\cite{XW-SM-BDOA:19,YH-ZM-JS:22}, where each agent accesses only a submatrix of the global coefficient matrix and transmits fixed-size state variables. This shifts the scalability challenge from time to space complexity but requires the total number of agents to scale with the solution dimension, which may not be practical. Alternatively, when the agent count is fixed, time complexity can be increased to reduce communication load. For instance,~\cite{JL-BDOA:20} proposes a scheme where agents cyclically transmit subsets of their estimates, while~\cite{SL-SM:25} updates and communicates only a portion of the solution per iteration using a nested-loop structure. The continuous-time approach in~\cite{LW-ZR-DY-GS:25} compresses agent states into scalars via time-varying transformations. Though effective under bandwidth constraints, these methods have limitations:~\cite{JL-BDOA:20} and~\cite{LW-ZR-DY-GS:25} require the LAE to have an exact solution, and~\cite{SL-SM:25} offers only practical convergence. Another related approach tackles finite data rate constraints using quantizers~\cite{JL-PY-GS-BDOA:20, LX-XY-JS-YS-KHJ-TY:25}, encoding messages as integer-valued vectors. However, these designs aim to enforce finite-bit messages, rather than fixed packet dimensions.}
	
	
	In this work, we propose a novel discrete-time scalable distributed least-squares (DT-SD-LS) algorithm for solving LAEs. Each agent has access to a subset of the LAE's coefficient matrix and the corresponding entries of the observation vector. To address the bandwidth constraint, we introduce a {\color{blue}cyclic} scheduling protocol that determines which portion of the solution needs to be transmitted at each iteration, similar to the approach in~\cite{JL-BDOA:20}. This scheduling results in a switching communication graph, a topic studied in~\cite{ROS-RMM:03c,LM:05,AN-AO:15} in the context of stability and consensus. However, these results are not directly applicable here due to the complex coupling between the time-varying communication network and the LS algorithm. Furthermore, we tackle the challenge of solving LAEs with time-varying observation vectors, a topic briefly mentioned in~\cite{BDOA-SM-ASM-UH:16}, and investigate whether exponential convergence rates with bounded tracking errors apply to switching algorithms like ours—an issue not fully explored in~\cite{XZ-QY-HW-WC-ZP-HF:23}. Our approach to handling time-varying observation vectors is conceptually similar to dynamic average consensus methods~\cite{MZ-SM:10-auto,SSK-BVS-JC-RAF-KML-SM:19}, though we focus on tracking a LS solution rather than an average state.
	
	The major contributions of this work are summarized below.
	\begin{enumerate}
		\item We first propose a novel continuous-time distributed least-squares (CT-D-LS) algorithm for LAEs. {\color{blue}Compared to other continuous-time algorithms in the literature, the CT-D-LS algorithm only needs to broadcast the ``guessed" solution, not the auxiliary vector, hence doubling the communication efficiency.} 
		\item The proposed DT-SD-LS algorithm is obtained by discretizing the CT-D-LS algorithm plus scheduling. under mild assumptions, 
		we prove that the agents' states driven by the DT-SD-LS algorithm will reach consensus, and their ``guessed" solution will converge to a LS solution of the LAE exponentially fast.
		\item When the LAE has multiple LS solutions, we characterize the LS solution to which the agents' ``guessed" solutions will converge to under the DT-SD-LS algorithm.
		\item When the observation vector is time-varying, we modify the DT-SD-LS algorithm in order for tracking a LS solution. We prove that practical convergence to a LS solution trajectory is guaranteed, with a tracking error bounded by a term proportional to the magnitude of the observation vector's single-step variation.
		\item {\color{blue}We apply the DT-SD-LS algorithm to medium-scale overdetermined and underdetermined LAE problems and compare it with other state-of-the-art distributed LS algorithms. The results show that the proposed DT-SD-LS algorithm is feasible, has competitive efficiency, and is scalable to larger problems.}
		\item Additionally, beyond the main theoretical results, we investigate the properties of the so-called semi-Hurwitz and semi-Schur matrices, which are of independent interest. 
	\end{enumerate}
	
	The rest of the paper is organized as follows. Section~\ref{sec:prelim} introduces the necessary background knowledge and provides the problem formulation. Section~\ref{sec:algo_full_state} presents a motivational continuous-time distributed algorithm. Section~\ref{sec:algo_scalabe} proposes the scalable distributed algorithm via scheduling and discusses its theoretical convergence guarantees. Section~\ref{sec:tracking} modifies the algorithm to address scenarios where the observation vector is time-varying. Section~\ref{sec:technical} contains the technical results developed for this work. Section~\ref{sec:simulation} validates our algorithms through two numerical examples. Finally, Section~\ref{sec:conclusion} concludes the paper.
	
	\section{Preliminaries}\label{sec:prelim}
	
	\paragraph*{Notations}
	We denote by $\N:=\{0,1,2,\cdots\}$ the set of non-negative integers,
	$\R$ the set of real numbers, {\color{blue}$\C$ the set of complex numbers, }
	{\color{blue}and $\cP:=\{1,\ldots,p\}, \cQ:=\{1,\ldots,q\}$ for some $p, q\in\N$}. In
	particular, $0_{n\times m}$, (resp.~$1_{n\times m}$) denotes the
	$n\times m$-dimensional zero matrix (resp.~all-ones matrix), while
	$I_n$ represents the $n\times n$ identity matrix. When the dimensions
	are clear from the context, we remove the subindices. A diagonal matrix composed by elements $a_1, \cdots, a_p$ is denoted by $\diag\left(a_1, \cdots, a_p\right)$, and a block diagonal matrix composed by matrices  $A_1,\cdots, A_p$ is denoted by $\diag\left(A_1,\cdots,A_p\right)$. The Kronecker product of two matrices $A, B$ is denoted by $A \otimes B$, and the transpose of matrix $A$ is denoted by $A^\top$. For any vector $x\in\R^n$ and matrix $A\in\R^{m\times n}$, let $\Vert x \Vert,\Vert A\Vert$ denote the 2-norm and induced norm, respectively. The range and
	kernel of a matrix $A\in\R^{m\times n}$ are denoted by
	$\ran(A),\ker(A)$, respectively. {\color{blue}For square matrix $A\in\R^{n\times n}$, let $\Lambda(A)$ be the spectrum of $A$, i.e., the set of all eigenvalues of $A$. For $\lambda\in\C$, let $\real(\lambda), \imag(\lambda)$ be its real and imaginary parts, respectively.}
	
	Let $\cU$ and $\cW$ be subspaces of $\cV$. We say that $\cV$ is the \emph{direct sum} of $\cU,\cW$, denoted as $\cV = \cU\oplus\cW$, if $\cV=\cU+\cW$ and $\cU\cap\cW=\{0\}$. For any $v\in\cU\oplus\cW$, we say that $u\in\cU$ is the \emph{oblique projection} of $v$ onto $\cU$ along $\cW$ if $v-u\in\cW$. Note that the oblique projection is unique, and when $v=u+w$ such that $u\in\cU$ and $w\in\cW$, $u$ is the oblique projection of $v$ onto $\cU$ along $\cW$ and 
	$w$ is the oblique projection of $v$ onto $\cW$ along $\cU$.
	
	In this work, we will frequently use the following two types of square matrices.
	
	\begin{definition}\label{def:semi-Hurwitz}
		A matrix $A\in\R^{n\times n}$ is \emph{semi-Hurwitz} if all its eigenvalues either have negative real parts, or are $0$ and non-defective (i.e., the geometric multiplicity and algebraic multiplicity of the eigenvalue $0$ are equal.)
	\end{definition}
	\begin{definition}\label{def:semi-Schur}
		A matrix $A\in\R^{n\times n}$ is \emph{semi-Schur} if all its eigenvalues either have magnitude strictly less than $1$, or are $1$ and non-defective.
	\end{definition}
	Note that a matrix being semi-Hurwitz or semi-Schur is stronger than being Lyapunov stable. Properties and applications of semi-Hurwitz and semi-Schur matrices will be studied in Section~\ref{subsec:semi-Hurwitz}.

	\paragraph*{Graph theory}
	An undirected graph $G=(\cP,E)$ consists of the vertex set $\cP$ and the edge set $E\subseteq\cP\times\cP$, such that
	$(i,j)\in E$ if and only if $(j,i)\in E$. A path is a sequence
	of vertices connected by edges, and the graph $G$ is
	\emph{connected} if there is a path between any pair of
	vertices. For any $i\in \cP$, the set of neighbors of $i$ is
	$\cN(i):=\{j\in \cP: (i,j)\in E\}$. A \emph{Laplacian matrix} $L=[L_{ij}]\in\R^{p\times p}$ of a graph $G$ is given by
	\begin{equation*}
		L_{ij}=\begin{cases}
			-1,&\mbox{if }(i,j)\in E, i\neq j,\\
			0,&\mbox{if }(i,j)\not\in E, i\neq j,\\
			-\sum_{k\neq i}L_{ik},&\mbox{if } i=j.
		\end{cases}
	\end{equation*}
	{\color{blue}For an undirected graph $G$, $L$ is symmetric and positive semi-definite. If $G$ is also connected, then $\ker(L)=\ran(1_{p\times 1})$.}

	\paragraph*{Problem formulation}
	A \emph{coefficient matrix} $A\in\R^{m\times n}$ and an \emph{observation vector} $b\in\R^{m}$ defines a \emph{linear algebraic equation} (LAE):
	\begin{equation}\label{LAE_0}
		Ax=b,
	\end{equation}
	in the unknowns $x\in\R^n$. Note that there may not exist $x$ so that \eqref{LAE_0} holds, especially in the case when $m>n$, i.e., when the problem is \emph{overdetermined}. Two types of solutions are considered in this work.
	
	\begin{definition}
		A vector $x^*\in\R^n$ is called an \emph{exact} solution to the problem \eqref{LAE_0} if $Ax^*=b$.
	\end{definition}
	
	\begin{definition}
		A vector $x^*\in\R^n$ is called a \emph{least squares} (LS) solution to the problem \eqref{LAE_0} if
		\begin{equation*}
			x^*\in\arg\min_{x\in\R^n}\Vert Ax-b\Vert.
		\end{equation*}
	\end{definition}
	Clearly, if $x$ is an exact solution, then it is also a LS solution.
	
	In this work, we study the problem of solving LAEs by a group of agents. specifically,
	
	\begin{problem}\label{prob:1}
		Find a LS solution to the LAE
		\begin{equation}\label{LAE}
			\begin{bmatrix}
				A_1\\\vdots\\A_p
			\end{bmatrix}x=\begin{bmatrix}
				b_1\\\vdots\\b_p
			\end{bmatrix},
		\end{equation}
		where $A_i\in\R^{m_i\times n}$, $b_i\in\R^{m_i}$, $\sum_{i=1}^{p}m_i=m$ by $p$ agents with a time-invariant connected communication graph $G=(\cP,E)$, subject to the following 3 constraints:
		\begin{enumerate}
			\item(Local information) The $i$-th agent only knows $A_i,b_i$.
			\item(Local communication) Each agent can only communicate with its neighbors defined by $G$.
			\item(Limited bandwidth) Each agent can only broadcast a packet $\zeta\in\R^{\bar n}$ with $\bar n< n$ per iteration of communication.	
		\end{enumerate}
	\end{problem}
	
	Our scalable distributed approach for solving Problem~\ref{prob:1} is conceptually illustrated in Fig.~\ref{fig:illustration}. For each $i\in\cP$, the $i$-th agent stores in memory its private data $A_i,b_i$, and a state vector $\xi_i=\begin{bmatrix}
		x_i\\v_i
	\end{bmatrix}\in\R^{N}$ for some $N>n$. The component $x_i\in\R^n$ represents the agent's estimate of a LS solution of \eqref{LAE}, and the component $v_i\in\R^{N-m}$ is an auxiliary vector which facilitates refining $x_i$. The state vector $\xi_i$ is updated at each iteration; however, since $N>\bar n$, the entire vector of $\xi_i$ cannot be broadcast to the neighboring agents at each iteration. Instead, a subset of at most $\bar n$ elements from $\xi_i$, denoted as a vector $\xi_i^{S(k)}\in\R^{n^{S(k)}_i}$ (where $n^{S(k)}_i\leq\bar n$), is selected and transmitted to neighbors at iteration $k$. Based on their own state vectors, and the partial state vectors received, each agent updates $\xi_i$ according to a rule of the form
	\begin{equation}\label{abstraction}
		\xi_i^+=f_i\Big(\xi_i,\{\xi_l^{S(k)}\}_{l\in\cN(i)}\Big).
	\end{equation}
	Our goal is to design the selection of elements $k\mapsto\xi_i^{S(k)}$, referred to as the \emph{scheduling} protocol, and design the maps $f_i:\R^{N+\sum_{l\in\cN(i)}n^{S(k)}_l}\to\R^N$, $i\in\cP$, referred to as the \emph{algorithm}, such that the components $x_i$'s eventually reach consensus and converge to a LS solution of the LAE~\eqref{LAE}.
	
	%
	
	\begin{figure}
		\centering
		\begin{tikzpicture}[scale=1.1]
	\foreach \i in {1,2,3}{
		\node[] at (\i*4-2.5,6.5) {Agent \i};
		\draw[fill=mycolor5] ((\i*4-4,5) rectangle +(3,1) node[pos=.5] {$A_\i$,$b_\i$};
		\draw[fill=mycolor2,fill opacity=0.6,text opacity=1] ((\i*4-4,5) rectangle +(3,-3) node[pos=.2] {$\xi_\i$};
		\draw [dashed] ((\i*4-3,2.5) rectangle +(1.5,1) node[pos=.5] (xi\i) {$\xi_\i^{S(k)}$};
		\draw[color=mycolor6,line width=3pt,-{Latex[width=8pt,length=5pt]}] (\i*4-2,2.2) to [out=330, in=210, looseness=8] node[color=black,below,rotate=10] {$\footnotesize\xi_\i^+=f_\i\Big(\xi_\i,\{\xi_l^{S(k)}\}_{l\in\mathcal N(\i)}\Big)$} (\i*4-3,2.2);
	}
	\draw[dotted, fill=black, fill opacity=0.1, text opacity=1] (-1,2.3) rectangle (13,4) node[pos=0.95,text width=1.2cm] {Comm. graph $G$};
	
	\draw [color=gray,line width=3pt,{Latex[width=8pt,length=5pt]}-{Latex[width=8pt,length=5pt]}] (xi1.east) -- (xi2.west);
	\draw [color=gray,line width=3pt,{Latex[width=8pt,length=5pt]}-{Latex[width=8pt,length=5pt]}] (xi2.east) -- (xi3.west);
	\draw [color=gray,line width=3pt,{Latex[width=8pt,length=5pt]}-] (xi1.west) -- +(-1.5,0);
	\draw [color=gray,line width=3pt,{Latex[width=8pt,length=5pt]}-] (xi3.east) -- +(1,0);
\end{tikzpicture}
		\caption{Illustration of a scalable distributed algorithm for solving Problem~\ref{prob:1}.}\label{fig:illustration} 
	\end{figure}
	
	\section{A continuous-time distributed least square algorithm}\label{sec:algo_full_state}
	
	We first introduce the CT-D-LS algorithm that can be used to find a LS solution of \eqref{LAE}, along with its convergence guarantees. It is important to note that the concept of ``iteration" does not apply to continuous-time algorithms, rendering the consideration of limited bandwidth irrelevant. Thus we do not require the CT-D-LS algorithm to be scalable. {\color{blue}Meanwhile, the CT-D-LS algorithm does not need to transmit any auxiliary vectors--an advantage over existing distributed approaches--while serving as a theoretical foundation for the subsequent development and analysis of the DT-SD-LS algorithm.}
	
	\subsection{Intuition and algorithm formulation}
	
	Let $L\in\R^{p\times p}$ be the Laplacian matrix of $G$, and let $\cN(i)$ be the set of neighbors of vertex $i$ in $G$. In addition, for each $i\in\cP$, we partition $A_i$ into $q\in\N$ sub-matrices:
	\begin{equation*}
		A_i:=\begin{bmatrix}
			A_{i1}&\cdots&A_{iq}
		\end{bmatrix},
	\end{equation*} 
	where $A_{ij}\in\R^{m_i\times n_j}$, $\sum_{j=1}^qn_j=n$ and $n_j\leq \bar n$ for all $j\in\cQ$. Let $G_{\textnormal{c}}=(\cQ,E_{\textnormal{c}})$ be an undirected connected graph over $q$ vertices, where each vertex represents an element of the partition. Let $L_{\textnormal{c}}\in\R^{q\times q}$ be the Laplacian matrix of $G_{\textnormal{c}}$, and let $\Nc(j)$ be the set of neighbors of vertex $j$ in $G_{\textnormal{c}}$.
	
	We now restructure the data $A_i, b_i$ so that they are more accessible by a scalable distributed algorithm. For all $i\in\cP, j\in\cQ$, let $x_{ij}\in\R^{n_j}$ be the $i$-th agent's guess for the $j$-th portion of the LS solution. In this way, each agent's guess of the full solution is $x_i=\begin{bmatrix}
		x_{i1}^\top& x_{i2}^\top&\ldots &x_{iq}^\top
	\end{bmatrix}^\top\in\R^n$. 
	
	In order to facilitate the algorithm design, we permute the elements of $x_1,\ldots,x_p$ and combine them into a $pn$ dimensional vector. To this end, we further denote
	\begin{equation*}
		\hat x_j:=\begin{bmatrix}
			x_{1j}\\\vdots\\x_{pj}
		\end{bmatrix}\in\R^{pn_j}\quad\forall j\in\cQ,
			\quad\hat x:=\begin{bmatrix}
				\hat x_1\\\vdots\\ \hat x_q
			\end{bmatrix}\in \R^{pn}.
		\end{equation*}
		Note that $\hat x_j$ consists of all agents' guesses of the $j$-th portion of a LS solution. Thus when consensus on the LS solution is reached, we should have $(L\otimes I_{n_j})\hat x_j=0$ for all $j\in\cQ$. Equivalently, $\hat L \hat x=0$, where
		\begin{equation*}
			\hat L:=\diag(L\otimes I_{n_1},\ldots,L\otimes I_{n_q})\in\R^{(pn)\times(pn)}.
		\end{equation*}
		Let $b_i=\sum_{j=1}^q b_{ij}$, where $b_{ij}$ can be arbitrarily chosen by the $i$-th agent. We can similarly define the augmented coefficient matrix and the augmented observation vector as follows. 
		\begin{align*}
			\hat A_j&:=\diag(A_{1j},\ldots,A_{pj})\in\R^{m\times(pn_j)}\quad\forall j\in\cQ,\\
			\hat A&:=\diag(\hat A_1,\ldots,\hat A_q)\in\R^{(qm)\times(pn)},
		\end{align*}
		\begin{equation*}
			\hat b_j:=\begin{bmatrix}
				b_{1j}\\\vdots\\b_{pj}
			\end{bmatrix}\in\R^m\quad\forall j\in\cQ,\quad \hat b:=\begin{bmatrix}
				\hat b_1\\\vdots\\\hat b_q
			\end{bmatrix}\in\R^{qm}.
		\end{equation*}
		Additionally, the augmented Laplacian matrix of $\cG_\textnormal{c}$ is defined by
		\begin{equation*}
			\Lc:=L_{\textnormal{c}}\otimes I_m\in\R^{(qm)\times(qm)}.
		\end{equation*}
		Note that since both $G,G_{\textnormal{c}}$ are undirected, $\hat L, \Lc$ are symmetric and positive semi-definite. Hence their square roots exist and are symmetric; they are denoted as $\hat L^{\frac{1}{2}}, \Lc^{\frac{1}{2}}$, respectively.
		
		The following result is adopted from~\cite{YH-ZM-JS:22}, which is fundamental for the design of our distributed algorithms.
		\begin{lemma}[{\cite[Lemma~3.1]{YH-ZM-JS:22}}]\label{lem:fundamental}
			Suppose both $G,G_c$ are connected. Then,
			\begin{equation*}
				x^*=\begin{bmatrix}
					x_1^*\\\vdots\\x_q^*
				\end{bmatrix}\in\R^n
			\end{equation*}
			is a LS solution of \eqref{LAE} if and only if there exists $\hat\omega^*\in\R^{qm}$ such that $(\hat x^*,\hat\omega^*)$ is an optimal solution to the problem
			\begin{subequations}\label{op_prob}
				\begin{align}
					\min_{\hat x\in \R^{pn}, \hat\omega\in \R^{qm}}\frac{1}{2}\Vert\hat A\hat x- \hat b-\Lc^{\frac{1}{2}}\hat\omega \Vert^2,\\
					\mbox{s.t. }\hat L\hat x=0,\label{op_prob_const}
				\end{align}
			\end{subequations}
			where
			\begin{equation}\label{hat_x^*&x^*}
				\hat x^*=\begin{bmatrix}
					\hat x^*_1\\\vdots\\\hat x^*_q
				\end{bmatrix},\quad \hat x^*_j=1_p\otimes x_j^*\quad\forall j\in\cQ.
			\end{equation}	
		\end{lemma}
		
		When there exists an optimizer $(\hat x, \hat\omega)$ such that the minimum of \eqref{op_prob} is $0$, solving \eqref{op_prob} is equivalent to solving an unconstrained convex optimization problem
		\begin{equation}\label{op_prob_2}
			\min_{\hat x\in \R^{pn}, \hat\omega\in \R^{qm}}\frac{1}{2}\Vert\hat A\hat x- \hat b-\Lc^{\frac{1}{2}}\hat\omega \Vert^2+\frac{k_P}{2}\hat x^\top \hat L \hat x,
		\end{equation}
		where $k_P>0$ is a tunable parameter. The problem~\eqref{op_prob_2} can be solved by the gradient flow:
		\begin{subequations}\label{poor_gradient_flow}
			\begin{align}
				\dot{\hat x}&=-\hat A^\top(\hat A\hat x-\hat b-\Lc^{\frac{1}{2}}\hat\omega)-k_P\hat L\hat x,\label{poor_gradient_flow_x}\\
				\dot{\hat \omega}&=\Lc^{\frac{1}{2}}(\hat A\hat x-\hat b-\Lc^{\frac{1}{2}}\hat\omega).
			\end{align}	
		\end{subequations}
		Nevertheless, when the LAE~\eqref{LAE} only has LS solution but no exact solution, the solutions of \eqref{op_prob} do not coincide with the solutions of \eqref{op_prob_2}.  In other words, the algorithm \eqref{poor_gradient_flow} will not give a solution of \eqref{op_prob} in this case. Inspired by the integrator method in \cite{XW-SM-BDOA:23}, we add an integrator term to \eqref{poor_gradient_flow_x}, aiming to enforce the achievement of the constraint \eqref{op_prob_const}:
		\begin{equation*}
			\dot{\hat x}=-\hat A^\top(\hat A\hat x-\hat b-\Lc^{\frac{1}{2}}\hat\omega)-k_P\hat L\hat x-k_I\int_0^t \hat L\hat x(s) ds,
		\end{equation*}
		where $k_I>0$ is another tunable parameter. 
		By defining $\hat z(t):=\int_0^t \hat L\hat x(s) ds$, the modified algorithm is given by
		
		\begin{subequations}\label{continuous-time_law_0}
			\begin{align}
				\dot{\hat x}&=-\hat A^\top(\hat A\hat x-\hat b-\Lc^{\frac{1}{2}}\hat\omega)-k_P\hat L\hat x-k_I\hat z,\label{continuous_time_law_0_x}\\
				\dot{\hat \omega}&=\Lc^{\frac{1}{2}}(\hat A\hat x-\hat b-\Lc^{\frac{1}{2}}\hat\omega),\label{continuous_time_law_0_omega}\\
				\dot{\hat z}&=\hat L\hat x,\label{continuous_time_law_0_z}
			\end{align}	
		\end{subequations}
		The algorithm \eqref{continuous-time_law_0} can be further simplified, by defining 
		\begin{equation}\label{def_y}
			\hat y:=\hat A\hat x-\hat b-\Lc^{\frac{1}{2}}\hat\omega.
		\end{equation}
		We have
		\begin{subequations}\label{continuous-time-law_1}
			\begin{align}
				\dot{\hat x}&=-\hat A^\top\hat y-k_P\hat L\hat x-k_I\hat z,\\
				\dot{\hat y}&=\hat A\dot{\hat x}-\Lc\hat y,\\
				\dot{\hat z}&=\hat L\hat x.
			\end{align}
		\end{subequations}
		Agent-wise and portion-wise,
		\begin{subequations}\label{continuous-time-law_1-componentwise}
			\begin{align}
				\dot x_{ij}&=-A_{ij}^\top y_{ij}-k_P\sum_{l\in\cN(i)}(x_{ij}-x_{lj})-k_Iz_{ij},\\
				\dot y_{ij}&=A_{ij}\dot x_{ij}-\sum_{\ell\in\Nc(j)}(y_{ij}-y_{i\ell}),\\
				\dot z_{ij}&=\sum_{l\in\cN(i)}(x_{ij}-x_{lj}),
			\end{align}
		\end{subequations}
		where
		\begin{align*}
			\hat y&:=\begin{bmatrix}
				\hat y_1\\\vdots \\\hat y_q
			\end{bmatrix},\quad \hat y_j=\begin{bmatrix}
				y_{1j}\\\vdots\\ y_{pj}
			\end{bmatrix},\quad y_{ij}\in\R^{m_i},\\
			\hat z&:=\begin{bmatrix}
				\hat z_1\\\vdots \\\hat z_q
			\end{bmatrix},\quad \hat z_j=\begin{bmatrix}
				z_{1j}\\\vdots\\ z_{pj}
			\end{bmatrix},\quad z_{ij}\in\R^{n_j},
		\end{align*}
		for all $i\in\cP,j\in\cQ$. This means in addition to the ``guessed" solution $x_i$, we also let the $i$-th agent control the auxiliary vectors $y_i:=\begin{bmatrix}
			y_{i1}\\\vdots\\y_{iq}
		\end{bmatrix}$, $z_i:=\begin{bmatrix}
			z_{i1}\\\vdots\\z_{iq}
		\end{bmatrix}.$
		It can be easily checked that the two assumptions of local information and local communication in Problem~\ref{prob:1} are satisfied. {\color{blue} The CT-D-LS algorithm is summarized in the following pseudo-code.}

		\begin{algorithm}[H]
			\caption{CT-D-LS}\label{alg:CT-D-LS}
			{\color{blue}	\begin{algorithmic}[1]
					\Require $k_I, k_P$.
					\State Initialize $x_{ij}(0)=0, y_{ij}(0)=-b_{ij}, z_{ij}(0)=0$ for all $i\in\cP, j\in\cQ$.
					\While{running, the $i$-th agent, $i\in\cP$}
					\State Broadcast $x_{ij}$, for all $j\in\cQ$ to its  neighbors.
					\State Update $x_{ij}, y_{ij}, z_{ij}$ for all $j\in\cQ$ according to the flows \eqref{continuous-time-law_1-componentwise}.
					\EndWhile
					\Ensure $x_i=\begin{bmatrix}
						x_{i1}\\\vdots\\x_{ip}
					\end{bmatrix}$, $i\in\cP$.
			\end{algorithmic}}
		\end{algorithm}
		{\color{blue}In the next subsection, we will show that the flows~\eqref{continuous-time_law_0} converge for any initial states of $\hat x,\hat \omega, \hat z$. However, for the CT-D-LS algorithm to converge, the initial states $x_{ij}(0), y_{ij}(0)$ must satisfy the transformation relationship \eqref{def_y} for some $\hat\omega(0)\in\R^{qm}$. By choosing $\hat\omega(0)=0$, the condition~\eqref{def_y} simplifies to local conditions that $y_{ij}(0)=A_{ij}x_{ij}(0)-b_{ij}$ for all $i\in\cP, j\in\cQ$. The specific choice in Step 1 of Algorithm~\ref{alg:CT-D-LS} represents one particularly convenient implementation of these valid initial states for the CT-D-LS algorithm.}
		
		{\color{blue}\begin{remark}[Comparison with other continuous-time distributed LS algorithms]\label{rem:CT_comparison}
				We compare Algorithm~\ref{alg:CT-D-LS} with the continuous-time distributed LS algorithms from the literature. 
				The works \cite{JW-NE:10}, \cite{BG-JC:14-tac} inspire the continuous-time flow
				\begin{subequations}\label{variety_2}
					\begin{align}
						\dot x_{i}&=-A_{i}^\top (A_{i}x_{i}-b_{i})-k_P\sum_{l\in\cN(i)}(x_{i}-x_{l})-\sum_{l\in\cN(i)}(z_{i}-z_{l}),\label{variety_2_1}\\
						\dot z_{i}&=\sum_{l\in\cN(i)}(x_{i}-x_{l}).
					\end{align}
				\end{subequations}
				for finding LS solutions of LAEs. Meanwhile, the works \cite{XW-JZ-SM-MJC:19,YL-CL-BDOA-GS:19} ignore the proportional diffusion term in \eqref{variety_2_1}, and rely on the update law
				\begin{subequations}\label{variety_3}
					\begin{align}
						\dot x_{i}&=-A_{i}^\top (A_{i}x_{i}-b_{i})-k_P\sum_{l\in\cN(i)}(z_{i}-z_{l}),\\
						\dot z_{ij}&=\sum_{l\in\cN(i)}(x_{i}-x_{l}).
					\end{align}
				\end{subequations}
				to establish their distributed LS algorithms. Additionally, the discrete-time distributed LS algorithm in \cite{TY-JG-JQ-XY-JW:20} can be viewed as a discretization of the continuous-time update law
				\begin{subequations}\label{variety_4}
					\begin{align}
						\dot x_{i}&=-\sum_{l\in\cN(i)}(x_{i}-x_{l})-k_Iz_{i},\\
						\dot z_{i}&=-\sum_{l\in\cN(i)}(z_{i}-z_{l})+A_{i}^\top A_{i}\dot x_{i}.
					\end{align}
				\end{subequations} 
				Finally, for our CT-D-LS algorithm, we note that if $G_c$ consists only a single vertex, then the auxiliary $\hat\omega$ becomes redundant and the update law~\eqref{continuous-time-law_1} reduces to
				\begin{subequations}\label{variety_1}
					\begin{align}
						\dot x_{i}&=-A_{i}^\top (A_{i}x_{i}-b_{i})-k_P\sum_{l\in\cN(i)}(x_{i}-x_{l})-k_Iz_{i},\\
						\dot z_{i}&=\sum_{l\in\cN(i)}(x_{i}-x_{l}).
					\end{align}
				\end{subequations}
				All these algorithms share a common structure, where $x_i$ represents the ``guessed" solution of the $i$-th agent and an auxiliary state variable $z_i$, of the same dimension as $x_i$, is used to facilitate convergence. While the dynamics of $x_i,z_i$ are similar across these laws, a significant distinction arises between \eqref{variety_1} and the others. Specifically, except for \eqref{variety_1}, the update laws in \eqref{variety_2}-\eqref{variety_4} require an additional diffusion term $\sum_{l\in\cN(i)}(z_{i}-z_{l})$. This means that the agents must exchange not only $x_l$ but also $z_l$ with their neighbors. Given the communication limitations, this extra transmission of $z_l$ halves the efficiency. Moreover, there is no straightforward way to balance communication and computation in these update laws, which makes them less scalable for large problems.
		\end{remark}}
		

		\subsection{Convergence guarantee}
		Here, we present a theoretical analysis for Algorithm~\ref{alg:CT-D-LS}, summarized in the following theorem. The technical lemmas used are provided in Section~\ref{sec:technical}.
		\begin{theorem}\label{thm:continuous_law}
			For any $k_P, k_I>0$, there exists a LS solution of \eqref{LAE} $x^*=\begin{bmatrix}
				(x_1^*)^\top&\ldots&(x_q^*)^\top
			\end{bmatrix}^\top$, $x_j^*\in\R^{n_j}$ such that for any $i\in\cP,j\in\cQ$, each $x_{ij}(t)$ obtained by Algorithm~\ref{alg:CT-D-LS} converges exponentially to $x_j^*$.
		\end{theorem}
		
		\begin{proof}
			Let $(\hat x^*, \hat \omega^*)\in\R^{pn}\times\R^{qm}$ be a solution to the problem~\eqref{op_prob}. According to the KKT conditions \cite[Chapter 5.5.3]{SB-LV:04}, there exists $\hat\lambda^*\in\R^{pn}$ such that
			\begin{subequations}\label{KKT}
				\begin{align}
					\hat A^\top\big(\hat A\hat x^*-\hat b-\Lc^{\frac{1}{2}}\hat\omega^*\big)+\hat L\hat\lambda^*&=0,\label{opt_dynamic_x}\\
					\Lc^{\frac{1}{2}}\big(\hat A\hat x^*-\hat b-\Lc^{\frac{1}{2}}\hat\omega^*\big)&=0,\label{opt_dynamic_omega}\\
					\hat L\hat x^*&=0.\label{opt_dynamic_lambda}
				\end{align}
			\end{subequations}
			Equivalently, these equations can be expressed as
			\begin{equation}\label{KKT_matrx_0}
				M\xi^*=F\hat b,
			\end{equation}
			where $\xi^*:=\begin{bmatrix}\hat x^*\\\hat\omega^*\\\frac{1}{\sqrt{k_I}}\hat L^{\frac{1}{2}}\hat\lambda^*
			\end{bmatrix}\in\R^{2pn+qm}$,
			\begin{equation}\label{algorithm_matrix_ct_ti}
				M:=\begin{bmatrix}
					-\hat A^\top \hat A-k_P\hat L& \hat A^\top \Lc^{\frac{1}{2}}&-\sqrt{k_I}\hat L^\frac{1}{2}\\
					\Lc^{\frac{1}{2}}\hat A& -\Lc&0\\
					\sqrt{k_I}\hat L^\frac{1}{2}&0&0
				\end{bmatrix},
			\end{equation}  
			and
			\begin{equation}\label{def:F}
				F:=\begin{bmatrix}
					-\hat A^\top\\-\Lc^{\frac{1}{2}}\\0
				\end{bmatrix}\in\R^{(2pn+qm)\times qm}.
			\end{equation}
			Meanwhile, let $\tilde z(t)=\int_0^t\hat L^{\frac{1}{2}}\hat x(s)ds$. 
			Define the error states by 
			\begin{equation}\label{error_states}
				e=\begin{bmatrix}
					e_x\\e_\omega\\e_z
				\end{bmatrix}:=\begin{bmatrix}
					\hat x-\hat x^*\\
					\hat \omega-\hat\omega^*\\
					\sqrt{k_I}\tilde z-\frac{1}{\sqrt{k_I}}\hat L^{\frac{1}{2}}\hat\lambda^*
				\end{bmatrix}=\begin{bmatrix}
					\hat x\\\hat\omega\\\sqrt{k_I}\tilde z
				\end{bmatrix}-\xi^*.
			\end{equation}
			The dynamics for $e_x$ can be computed via \eqref{continuous_time_law_0_x}, \eqref{opt_dynamic_x}, \eqref{opt_dynamic_lambda}
			\begin{align*}
				\dot e_x&=\dot{\hat x}\\
				&=-\hat A^\top\big(\hat A(e_x+\hat x^*)-\hat b-\Lc^{\frac{1}{2}}(e_\omega+\hat\omega^*)\big)-k_P\hat L(e_x+\hat x^*)\\
				&\quad-k_I\hat L^{\frac{1}{2}}\big(\frac{1}{\sqrt{k_I}}e_z+\frac{1}{k_I}\hat L^{\frac{1}{2}}\hat\lambda^*\big)\\
				&=-\hat A^\top \hat Ae_x+\hat A^\top \Lc^{\frac{1}{2}}e_\omega-k_P\hat L e_x-\sqrt{k_I}\hat L^{\frac{1}{2}}e_z,
			\end{align*}
			where \eqref{opt_dynamic_x}, \eqref{opt_dynamic_lambda}, and the fact that $\hat z=\hat L^{\frac{1}{2}}\tilde z=\hat L^{\frac{1}{2}}\big(\frac{1}{\sqrt{k_I}}e_z+\frac{1}{k_I}\hat L^{\frac{1}{2}}\hat\lambda^*\big)$ are used. Similarly, from \eqref{continuous_time_law_0_omega} and \eqref{opt_dynamic_omega},
			\begin{align*}
				\dot e_\omega&=\dot{\hat\omega}\\
				&=\Lc^{\frac{1}{2}}\big(\hat A(e_x+x^*)-\Lc^{\frac{1}{2}}(e_\omega+\omega^*)-\hat b\big)\\
				&=\Lc^{\frac{1}{2}} \hat Ae_x- \Lc e_\omega,
			\end{align*}
			and, by using \eqref{continuous_time_law_0_z}, \eqref{opt_dynamic_lambda}, and the fact that $\ker(\hat L)=\ker(\hat L^{\frac{1}{2}})$,
			\begin{equation*}
				\dot e_z=\sqrt{k_I}\dot {\tilde z}=\sqrt{k_I}\hat L^{\frac{1}{2}}\hat x=\sqrt{k_I}\hat L^{\frac{1}{2}}(e_x+\hat x^*)=\sqrt{k_I}\hat L e_x.
			\end{equation*} 
				Thus, the error dynamics is governed by the continuous-time \emph{linear time-invariant} (LTI) system
				\begin{equation}\label{LTI_e}
					\dot e=Me,
				\end{equation}
				where recall that the matrix $M$ is given in \eqref{algorithm_matrix_ct_ti}. 
					
					With $\Theta=\begin{bmatrix}
						\hat A\\-\Lc^{\frac{1}{2}}
					\end{bmatrix}$, $\Xi=\begin{bmatrix}
						\hat L^{\frac{1}{2}}\\0
					\end{bmatrix}$, $a=k_p$ and $b=\sqrt{k_I}$, it follows from Lemma~\ref{lem:M} in Section~\ref{sec:technical} that $M$ is semi-Hurwitz. Let $\tilde e=\begin{bmatrix}
						\tilde e_x\\\tilde e_\omega\\\tilde e_z
					\end{bmatrix}$ be the oblique projection of $e(0)=-\xi^*=-\begin{bmatrix}
						\hat x^*\\\hat\omega^*\\\frac{1}{\sqrt{k_I}}\hat L^{\frac{1}{2}}\lambda^*
					\end{bmatrix}$ onto $\ker(M)$ along $\ran(M)$. Corollary~\ref{cor:semi-Hurwitz} in Section~\ref{sec:technical} applied to the dynamics \eqref{LTI_e} implies that $e(t)$ converges to $\tilde e$ exponentially fast. In addition, the fact that $\tilde e\in\ker(M)$, when combined with \eqref{KKT_matrx_0} and \eqref{error_states}, implies
					\begin{equation}
						M\lim_{t\to\infty}\begin{bmatrix}
							\hat x(t)\\\hat\omega(t)\\\sqrt{k_I}\tilde z(t)
						\end{bmatrix}=M\lim_{t\to\infty} (e(t)+\xi^*)=M\tilde e+M\xi^*=F\hat b,
					\end{equation}
					and the convergence is exponentially fast.
					Note that the problem~\eqref{op_prob} is convex so that the KKT conditions are also sufficient for optimality. As a result, the pair $\lim_{t\to\infty}(\hat x(t), \hat\omega(t))$ is an optimal solution to the problem~\eqref{op_prob}, with an adjoint state $\hat\lambda=k_I\int_0^\infty\hat x(s)ds\in\R^{pn}$. Lastly, by appealing to Lemma~\ref{lem:fundamental}, $\hat x$ converges to a LS solution $x^*$ of~\eqref{LAE} in the sense of \eqref{hat_x^*&x^*} exponentially fast.
				\end{proof}
				
				As reflected by the proof of Theorem~\ref{thm:continuous_law}, the matrix \eqref{algorithm_matrix_ct_ti} plays an important role in the convergence analysis. This matrix is called the \emph{algorithm matrix}. Based on the algorithm matrix $M$, we can precisely characterize the LS solution found by our CT-D-LS algorithm. This is given by the following proposition.
				
				\begin{proposition}\label{prop:x^*}
					There is a unique $\zeta^*\in\ran(M)$ which is also an exact solution of the LAE~\eqref{KKT_matrx_0}. Moreover, the LS solution $x^*$ of \eqref{LAE} in Theorem~\ref{thm:continuous_law} found by Algorithm~\ref{alg:CT-D-LS} is given by $\hat x^*=\begin{bmatrix}
						I_{np}&0_{(np)\times(2np)}
					\end{bmatrix}\zeta^*$ through \eqref{hat_x^*&x^*}.
				\end{proposition}
				\begin{proof}
					Because $M$ is semi-Hurwitz, it follows from Lemma~\ref{lem:direct_sum} in Section~\ref{sec:technical} that $\ran(M)\oplus\ker(M)=\R^{2pn+qm}$. Because the optimization problem~\eqref{op_prob} is feasible, exact solutions to the KKT condition~\eqref{KKT_matrx_0} exist. Let $\xi\in\R^{2pn+qm}$ be one of the exact solutions, and let $\xi_a$ be the oblique projection of $\xi$ onto $\ran(M)$ along $\ker(M)$. Since $\xi-\xi_a\in\ker(M)$, $\xi_a$ is also an exact solution to \eqref{KKT_matrx_0}. This proves that \eqref{KKT_matrx_0} has exact solutions in $\ran(M)$. Now, suppose both $\xi,\zeta\in\ran(M)$ are exact solutions of \eqref{KKT_matrx_0}. Then $\xi-\zeta\in\ran(M)\cap\ker(M)=\{0\}$ so the uniqueness of the solution is proven.
					
					Let $\xi^*$ be an exact solution of \eqref{KKT_matrx_0}, not necessarily $\xi^*\in\ran(M)$. It is shown in the proof of Theorem~\ref{thm:continuous_law} that $\hat x(t)$ converges to $\begin{bmatrix}
						I_{np}&0_{(np)\times(2np)}
					\end{bmatrix}(\xi^*+\tilde e)$, where recall that $\tilde e$ is the oblique projection of $-\xi^*$ onto $\ker(M)$ along $\ran(M)$. Hence $\xi^*+\tilde e\in\ran(M)$ and it is also an exact solution to \eqref{KKT_matrx_0}. By uniqueness of the solution, we must have $\xi^*+\tilde e=\zeta^*$, which completes the proof.
				\end{proof}
				
				{\color{blue}\begin{remark}[On selection of the tunable parameters]
						As stated in Theorem~\ref{thm:continuous_law}, convergence is guaranteed for any positive values of the tunable parameters $k_P$ and $k_I$. To achieve the fastest convergence rate, we consider the following optimization problem:
						\begin{equation}\label{EVP}
							\min_{k_I,k_P>0}\max_{\lambda\in\Lambda(M)\backslash\{0\}}\real(\lambda).
						\end{equation}
						Since the matrix $M$ is affine in the parameters $k_P$ and $\sqrt{k_I}$, the problem in \eqref{EVP} constitutes an eigenvalue optimization problem that can be solved via matrix inequalities~\cite{SB-LEG-EF-VB:94}. However, solving it requires centralized knowledge of the matrices $\hat A$ and $\hat L$, which is not available in our distributed setting. Moreover, the optimal choice of $k_P$ and $k_I$ for the continuous-time algorithm may not yield the fastest convergence in the discrete-time case due to discretization effects. A thorough investigation of optimal parameter selection for discrete-time performance is left for future work.
				\end{remark}}
				
				\section{A discrete-time scalable distributed least square algorithm via scheduling}\label{sec:algo_scalabe}
				
				Inspired by the CT-D-LS algorithm, we now establish the DT-SD-LS algorithm for solving LAEs. This development involves first discretizing the continuous-time algorithm and then incorporating scheduling to address the constraint of limited bandwidth. 
				
				\subsection{Discretization and scheduling}
				
				First of all, the continuous flows \eqref{continuous-time-law_1} can be implemented in a discrete-time manner, with a fixed step size $\alpha>0$:
				
				\begin{subequations}\label{discrete-time-law_1-componentwise}
					\begin{align}
						x_{ij}(k+1)&=x_{ij}(k)-\alpha\Big( A_{ij}^\top y_{ij}(k)\nonumber\\
						&\quad+k_P\sum_{l\in\cN(i)}(x_{ij}(k)-x_{lj}(k))+k_Iz_{ij}(k)\Big),\label{discrete-time-law_1-componentwise_x}\\
						y_{ij}(k+1)&=y_{ij}(k)+A_{ij}(x_{ij}(k+1)- x_{ij}(k))\nonumber\\
						&\quad-\alpha\sum_{\ell\in\Nc(j)}(y_{ij}(k)-y_{i\ell}(k)),\label{discrete-time-law_1-componentwise_y}\\
						z_{ij}(k+1)&=z_{ij}(k)+\alpha\sum_{l\in\cN(i)}(x_{ij}(k)-x_{lj}(k)).\label{discrete-time-law_1-componentwise_z}
					\end{align}
				\end{subequations}

				We have the following convergence result for this discrete-time algorithm:
				\begin{corollary}\label{cor:discrete_0}
					For sufficiently small $\alpha>0$, there exists a LS solution of \eqref{LAE} $x^*=\begin{bmatrix}
						(x_1^*)^\top&\ldots&(x_q^*)^\top
					\end{bmatrix}^\top$, $x_j^*\in\R^{n_j}$ such that for any $i\in\cP,j\in\cQ$, each $x_{ij}(k)$ of the algorithm~\eqref{discrete-time-law_1-componentwise} with initial states $x_{ij}(0)=0$, $y_{ij}(0)=-b_{ij}$ and $z_{ij}=0$converges  exponentially to $x_j^*$.
				\end{corollary}
				We do not prove Corollary~\ref{cor:discrete_0} here, since the update law~\eqref{discrete-time-law_1-componentwise} is a special case of the DT-SD-LS algorithm,  which will be discussed subsequently. Note that, similar to the CT-D-LS algorithm, only the states $x_{ij}$ need to be broadcast to the neighbors. However, since the set of neighbors $\cN(i)$ is independent of the portion index $j$, all portions, i.e., the entire guess of a solution, needs to be broadcast at each iteration. This requirement violates our limited bandwidth assumption and thus makes the algorithm non-scalable for large $n$. 
				
				{\color{blue}To address this limitation, we propose that only the portions $x_{1j},\ldots,x_{pj}$ for a single $j\in\cQ$ to be transmitted between the agents according to the communication graph $G$ at each iteration. For simplicity, we further implement a cyclic scheduling protocol; that is, we only transmit the portions $x_{1j},\ldots,x_{pj}$, where $j=(k\bmod q)+1$  at iteration $k$. Referring to the sketch in Fig.~\ref{fig:illustration} and the abstraction~\eqref{abstraction}, we essentially have
					\begin{equation*}
						\xi_i:=\begin{bmatrix}
							x_i\\y_i\\z_i
						\end{bmatrix},\quad x_i=\begin{bmatrix}
							x_{i1}\\\vdots\\x_{iq}
						\end{bmatrix},\quad y_i=\begin{bmatrix}
							y_{i1}\\\vdots\\y_{iq}
						\end{bmatrix},\quad z_i=\begin{bmatrix}
							z_{i1}\\\vdots\\z_{iq}
						\end{bmatrix},
					\end{equation*}
					as the local state vector, and
					\begin{equation*}
						\xi_i^{S(k)}=x_{ij},\mbox{ where }j=(k\bmod q)+1
					\end{equation*}
					as the message of bounded dimension that will be transmitted in the network. Because $x_{ij}\in\R^{n_j}$ and $n_j\leq \bar n$ for all $j\in\cQ$, this periodic scheduling protocol fulfills the limited bandwidth constraint. }
				
				{\color{blue}In terms of computation, at iteration $k$, we update $x_{ij}, y_{ij}$ and $z_{ij}$ according to \eqref{discrete-time-law_1-componentwise} only for $j=(k\bmod q)+1$. For the other portions, because $x_{lj}$'s are not received from the neighbors, we nullify the diffusion terms in \eqref{discrete-time-law_1-componentwise_x} and \eqref{discrete-time-law_1-componentwise_z}, and update $x_{ij}$ and $z_{ij}$ via
					\begin{align}
						x_{ij}(k+1)&=x_{ij}(k)-\alpha\Big( A_{ij}^\top y_{ij}(k)+k_Iz_{ij}(k)\Big),\label{discrete-time-law_1'-componentwise_x}\\
						z_{ij}(k+1)&=z_{ij}(k).\label{discrete-time-law_1'-componentwise_z}
					\end{align}
					Note that for those portions, their update law does not require any information from neighboring agents. Hence the computation is compatible with the communication protocol.
					Our DT-SD-LS algorithm is summarized in the following pseudo-code.}
				\begin{algorithm}[H]
					\caption{DT-SD-LS}\label{alg:DT-SD-LS}
					{\color{blue}	\begin{algorithmic}[1]
							\Require $\alpha,k_I, k_P$.
							\State Initialize $x_{ij}(0)=0, y_{ij}(0)=-b_{ij}, z_{ij}(0)=0$ for all $i\in\cP, j\in\cQ$.
							\For{$k=0,1,\ldots,k_{\max}-1$, the $i$-th agent, $i\in\cP$}
							\For{$j=1,\ldots,q$}
							\If{$j=(k\bmod q)+1$}
							\State Broadcast $x_{ij}(k)$ to the $\ell$-th agent for all $\ell\in\cN(i)$.
							\State Update $x_{ij}(k+1), y_{ij}(k+1), z_{ij}(k+1)$ for all $j\in\cQ$ according to \eqref{discrete-time-law_1-componentwise}.
							\Else
							\State Do not broadcast $x_{ij}(k)$.
							\State Update $x_{ij}(k+1), y_{ij}(k+1), z_{ij}(k+1)$ for all $j\in\cQ$ according to \eqref{discrete-time-law_1'-componentwise_x}, \eqref{discrete-time-law_1-componentwise_y} and \eqref{discrete-time-law_1'-componentwise_z}.
							\EndIf
							\EndFor
							\EndFor
							\Ensure $x_i=\begin{bmatrix}
								x_{i1}\\\vdots\\x_{ip}
							\end{bmatrix}$, $i\in\cP$.
					\end{algorithmic}}
				\end{algorithm}
				
				We also remark that the periodically-changing update-law made by \eqref{discrete-time-law_1-componentwise}, \eqref{discrete-time-law_1'-componentwise_x}, \eqref{discrete-time-law_1'-componentwise_z} can be written as a unified time-varying update-law
				\begin{subequations}\label{discrete-time-law_2-componentwise}
					\begin{align}
						x_{ij}(k+1)&=x_{ij}(k)-\alpha\Big( A_{ij}^\top y_{ij}(k)\nonumber\\
						&\quad+k_P\hspace{-5pt}\sum_{l\in\cN_j(k,i)}\hspace{-5pt}(x_{ij}(k)-x_{lj}(k))+k_Iz_{ij}(k)\Big),\label{discrete-time-law_2-componentwise_x}\\
						y_{ij}(k+1)&=y_{ij}(k)+A_{ij}(x_{ij}(k+1)- x_{ij}(k))\nonumber\\
						&\quad-\alpha\sum_{\ell\in\Nc(j)}(y_{ij}(k)-y_{i\ell}(k)),\label{discrete-time-law_2-componentwise_y}\\
						z_{ij}(k+1)&=z_{ij}(k)+\alpha\hspace{-5pt}\sum_{l\in\cN_j(k,i)}\hspace{-5pt}(x_{ij}(k)-x_{lj}(k)),\label{discrete-time-law_2-componentwise_z}
					\end{align}
				\end{subequations}
				where 
				\begin{equation*}
					\cN_j(k,i):=\begin{cases}
						\cN(i)&\mbox{ if }j=(k\bmod q)+1,\\
						\emptyset&\mbox{ otherwise}.
					\end{cases}
				\end{equation*}
				
				{\color{blue}\begin{remark}[Comparison with another discrete-time scalable distributed LS algorithm]\label{rem:DT_comparison}
						Here we compare Algorithm~\ref{alg:CT-D-LS} and the algorithm consisting of a double-layered network in \cite{YH-ZM-JS:22}. Algorithm~\ref{alg:CT-D-LS} is developed from an integral method, whereas algorithm~\eqref{their_algorithm} is based on a primal-dual approach such that its update law is given by
						\begin{subequations}\label{their_algorithm}
							\begin{align}
								x_{ij}^+&=x_{ij}-\alpha A_{ij}^\top y_{ij}-\beta \hspace{-5pt}\sum_{l\in\cN(i)}\hspace{-2pt}(x_{ij}-x_{lj})-\beta\hspace{-5pt} \sum_{l\in\cN(i)}\hspace{-2pt}(z_{ij}-z_{lj}),\\
								y_{ij}^+&=y_{ij}+ A_{ij} (x_{ij}^+-x_{ij})-\alpha\sum_{\ell\in\Nc(j)}(y_{ij}-y_{i\ell}),\\
								z_{ij}^+&=z_{ij}+\beta\sum_{l\in\cN(i)}(x_{ij}-x_{lj}).
							\end{align}
						\end{subequations}
						In the double-layered network architecture, the $ij$-th aggregator must broadcast $x_{ij}$ and $z_{ij}$ to its neighbors in the first-layer graph, and $y_{ij}$ to its neighbors in the second-layer graph. Although the size of the communication packets is independent of the solution dimension, this scalability comes at the cost of increasing the total number of agents in the network.
						Moreover, while update laws~\eqref{discrete-time-law_1-componentwise} and~\eqref{their_algorithm} appear structurally similar, they differ in a critical way as noted in Remark~\ref{rem:CT_comparison}: our update law~\eqref{discrete-time-law_1-componentwise} does not require the transmission of $z_{lj}$ from neighboring agents. As a result, its communication efficiency is effectively twice that of~\eqref{their_algorithm}. 
				\end{remark}}
				
				
				\subsection{Convergence guarantee}
				Since the DT-SD-LS algorithm is discretized from the CT-D-LS algorithm, the step size needs to be sufficiently small in order to guarantee convergence. In addition, the DT-SD-LS algorithm via scheduling requires the matrix
				\begin{equation}\label{def:tilde_M}
					\tilde M:=\begin{bmatrix}
						-\hat A^\top \hat A-\frac{k_P}{q}\hat L& \hat A^\top \Lc^{\frac{1}{2}}&-\sqrt{\frac{k_I}{q}}\hat L^\frac{1}{2}\\
						\Lc^{\frac{1}{2}}\hat A& -\Lc&0\\
						\sqrt{\frac{k_I}{q}}\hat L^\frac{1}{2}&0&0
					\end{bmatrix}
				\end{equation}
				to be diagonalizable. This is summarized in the following theorem.
				\begin{theorem}\label{thm:discrete_1}
					Suppose that the matrix $\tilde M$ defined in \eqref{def:tilde_M} is diagonalizable for some $k_P,k_I$. For sufficiently small $\alpha>0$, there exists a LS solution of \eqref{LAE} $x^*=\begin{bmatrix}
						(x_1^*)^\top&\ldots&(x_q^*)^\top
					\end{bmatrix}^\top$, $x_j^*\in\R^{n_j}$ such that for any $i\in\cP,j\in\cQ$, each $x_{ij}(k)$ obtained by Algorithm~\ref{alg:DT-SD-LS} converges exponentially to $x_j^*$.
				\end{theorem}
				
				Compared with~\eqref{algorithm_matrix_ct_ti}, the matrix $\tilde M$ can also be treated as an algorithm matrix, by scaling both tunable parameters $k_P,k_I$ by a factor of $\frac{1}{q}$. {\color{blue}Meanwhile, we note that the assumption of diagonalizability for $\tilde M$
					cannot be checked a priori, as it depends on $\hat A$, which is centralized. However, diagonalizability is a generic property of matrices. In all our simulations with randomly generated $A$ matrices, this assumption has always been satisfied.}
				
				
				Before presenting the proof of Theorem~\ref{thm:discrete_1}, we provide an intuitive discussion on the convergence of the CT-SD-LSA, inspired by the proof of Theorem~\ref{thm:continuous_law}. We first define the error states and derive the discrete-time error dynamics in a form analogous to the continuous-time case. Due to the periodic scheduling, these dynamics can be “averagely approximated” by a discrete-time LTI system with a semi-Schur system matrix, with the approximation improving as $\alpha$ decreases. In Theorem~\ref{thm:continuous_law}, we showed that convergence of a continuous-time LTI system with a semi-Hurwitz matrix leads to convergence of $x_{ij}$ to a LS solution of~\eqref{LAE}. Here, we aim to establish an analogous result for the discrete-time case, showing that convergence of the semi-Schur LTI system implies convergence of the actual dynamics, using technical tools developed in Section~\ref{sec:technical}.
				
				\begin{proof}
					Similar to the algorithm \eqref{continuous-time_law_0}, we consider the following discrete-time dynamics
					\begin{subequations}\label{discrete_time_law_2}
						\begin{align}
							\hat x(k+1)&=\hat x(k)-\alpha\Big(\hat A^\top(\hat A\hat x(k)-\hat b-\Lc^{\frac{1}{2}}\hat\omega(k))\nonumber\\
							&\quad+k_P\hat L_\s(k)\hat x(k)+k_I\hat z(k)\Big),\label{discrete_time_law_2_x}\\
							\hat \omega(k+1)&=\omega(k)+\alpha\Lc^{\frac{1}{2}}(\hat A\hat x(k)-\hat b-\Lc^{\frac{1}{2}}\hat\omega(k)),\label{discrete_time_law_2_omega}\\
							\hat z(k+1)&=\hat z(k)+\alpha\hat L_\s(k)\hat x(k),\label{discrete_time_law_2_z}
						\end{align}
					\end{subequations}
					where $\hat L_\s(k)=\hat L_{(k\bmod q)+1}$ and
					\begin{multline}\label{def:L_i}
						\hat L_j=\diag\big(0_{(p\sum_{k=1}^{j-1}n_k)\times(p\sum_{k=1}^{j-1}n_k)},L\otimes I_{n_j},\\
						0_{(p\sum_{k=j+1}^qn_k)\times(p\sum_{k=j+1}^qn_k)}\big).
					\end{multline}
					In other words, $\hat L_j$ is constructed from $\hat L$ by converting all the block diagonals to be $0$, except for the $j$-th block diagonal. Then, \eqref{discrete-time-law_2-componentwise} is exactly \eqref{discrete_time_law_2} written in the agent-wise and portion-wise form, with the change of variables~\eqref{def_y}. Moreover, we have the following observations based on \eqref{def:L_i}:
					\begin{equation}\label{ob_product}
						\hat L_j=\hat L^{\frac{1}{2}}\hat L_j^{\frac{1}{2}},\quad\forall j\in\cQ,
					\end{equation}
					\begin{equation}\label{ob_kernel_inclusion}
						\ker(\hat L)\subseteq\ker(\hat L_j),\quad\forall j\in\cQ,
					\end{equation}
					\begin{equation}\label{ob_image_inclusion}
						\ran(\hat L)\supseteq\ran(\hat L_j),\quad\forall j\in\cQ,
					\end{equation}
					and
					\begin{equation}\label{ob_summation}
						\hat L=\sum_{j=1}^q\hat L_j.
					\end{equation}
					
					Let $(\hat x^*, \hat \omega^*)\in\R^{pn}\times\R^{pn}$ be a solution to the problem~\eqref{op_prob}. Then the KKT condition \eqref{KKT} still holds with some $\hat\lambda^*\in\R^{pn}$.
					Let $\tilde z:\N\to\R^{pn}$ be defined by
					\begin{equation}\label{def:tilde_z}
						\tilde z(0)=0,\quad \tilde z(k)=\alpha\sum_{i=0}^{k-1}\hat L_\s^{\frac{1}{2}}(i)\hat x(i)\quad\forall k>0.
					\end{equation}
					It then follows from \eqref{discrete_time_law_2_z}, \eqref{ob_product}, and the initial state $\hat z(0)=0$ that
					$\hat z=\hat L^{\frac{1}{2}}\tilde z$.
					Define the error states by \eqref{error_states}. 
					The dynamics for $e_x$ can be computed via \eqref{opt_dynamic_x}, \eqref{opt_dynamic_lambda}, \eqref{discrete_time_law_2_x}, \eqref{ob_kernel_inclusion}
					\begin{align*}
						e_x(k+1)&=\hat x(k+1)-\hat x^*\\
						&=\hat x(k)-\hat x^*\\
						&\quad-\alpha\Big(\hat A^\top\big(\hat A(e_x(k)+\hat x^*)-\hat b-\Lc^{\frac{1}{2}}(e_\omega(k)+\hat\omega^*)\big)\\
						&\quad+k_P\hat L_\s(k)(e_x(k)+\hat x^*)\\
						&\quad+k_I\hat L^{\frac{1}{2}}\big(\frac{1}{\sqrt{k_I}}e_z(k)+\frac{1}{k_I}\hat L^{\frac{1}{2}}\lambda^*\big)\Big)\\
						&=e_x(k) -\alpha\big(\hat A^\top \hat Ae_x(k)-\hat A^\top \Lc^{\frac{1}{2}}e_\omega(k)\\
						&\quad+k_P\hat L_\s(k) e_x(k)+\sqrt{k_I}\hat L^{\frac{1}{2}}e_z(k)\big),
					\end{align*}
					where the fact that $\hat z=\hat L^{\frac{1}{2}}\tilde z=\hat L^{\frac{1}{2}}\big(\frac{1}{\sqrt{k_I}}e_z+\frac{1}{k_I}\hat L^{\frac{1}{2}}\lambda^*\big)$ is also used. Meanwhile, by using \eqref{opt_dynamic_omega} and \eqref{discrete_time_law_2_omega}, 
					\begin{align*}
						e_\omega(k+1)&=\hat\omega(k+1)-\hat\omega^*\\
						&=\hat\omega(k)-\hat\omega^*\\
						&\quad+\alpha\Lc^{\frac{1}{2}}\big(\hat A(e_x(k)+x^*)-\Lc^{\frac{1}{2}}(e_\omega(k)+\omega^*)-\hat b\big)\\
						&=e_\omega(k)+\alpha\big(\Lc^{\frac{1}{2}} \hat Ae_x(k)- \Lc e_\omega(k)\big),
					\end{align*}
					and by using \eqref{opt_dynamic_lambda}, \eqref{discrete_time_law_2_z}, and \eqref{ob_kernel_inclusion},
					\begin{align*}
						e_z(k+1)&=\sqrt{k_I}\tilde z(k+1)-\frac{1}{\sqrt{k_I}}\hat L^{\frac{1}{2}}\lambda^*\\
						&=\sqrt{k_I}\big(\tilde z(k)+\alpha\hat L_\s^{\frac{1}{2}}(k)\hat x(k)\big)-\frac{1}{\sqrt{k_I}}\hat L^{\frac{1}{2}}\lambda^*\\
						&=e_z(k)+\alpha\sqrt{k_I}\hat L_\s^{\frac{1}{2}}(k)(e_x+\hat x^*)\\
						&=e_z(k)+\alpha\sqrt{k_I}\hat L_s (k)e_x.
					\end{align*} 
					We conclude the compact form for the error dynamics that \begin{equation}\label{discrete_time_LTV}
						e(k+1)=\big(I+\alpha M_\s(k)\big)e(k),
					\end{equation}
					where 
					\begin{equation}\label{algorithm_matrix_dt_tv}
						M_\s(k):=\begin{bmatrix}
							-\hat A^\top \hat A-k_P\hat L_\s(k)& \hat A^\top \Lc^{\frac{1}{2}}&-\sqrt{k_I}\hat L^\frac{1}{2}\\
							\Lc^{\frac{1}{2}}\hat A& -\Lc&0\\
							\sqrt{k_I}\hat L^\frac{1}{2}_s(k)&0&0
						\end{bmatrix}.
					\end{equation}  
					Let
					\begin{multline}\label{def:bar_M}
						M_\ave:=\frac{1}{q}\sum_{k=0}^{q-1}M_\s(k)\\
						=\begin{bmatrix}
							-\hat A^\top \hat A-\frac{k_P}{q}\hat L& \hat A^\top \Lc^{\frac{1}{2}}&-\sqrt{k_I}\hat L^\frac{1}{2}\\
							\Lc^{\frac{1}{2}}\hat A& -\Lc&0\\
							\frac{\sqrt{k_I}}{q}\hat L^\frac{1}{2}&0&0
						\end{bmatrix},
					\end{multline}
					where the observation~\eqref{ob_summation} is used. Comparing $M_\ave$ and the algorithm matrix $M$ in \eqref{algorithm_matrix_ct_ti}, we have $\ker(M_\ave)=\ker(M)$. Additionally, $M_\ave=T^{-1}\tilde MT$, where $T=\diag(I,I,\sqrt q I)$ and $\tilde M$ is defined by \eqref{def:tilde_M}. With $\Theta=\begin{bmatrix}
						\hat A\\-\Lc^{\frac{1}{2}}
					\end{bmatrix}$, $\Xi=\begin{bmatrix}
						\hat L^{\frac{1}{2}}\\0
					\end{bmatrix}$, $a=\frac{k_p}{q}$ and $b=\sqrt{\frac{k_I}{q}}$, it follows from Lemma~\ref{lem:M} in Section~\ref{sec:technical} that the matrix $\tilde M$ is semi-Hurwitz. It is also diagonalizable by the assumption in Theorem~\ref{thm:discrete_1}. Hence, $M_\ave$ is similar to a semi-Hurwitz and diagonalizable matrix and it is also semi-Hurwitz and diagonalizable. Moreover, it is implied by the observation \eqref{ob_kernel_inclusion} that $\ker(M_\ave)\subset\ker(M_\s(j))$ for all $j=0,1,\ldots,q-1$. We can thus conclude from Lemma~\ref{lem:eigen_pert} in Section~\ref{sec:technical} that when $\alpha$ is sufficiently small, the matrix
					\begin{equation}\label{def:Phi}
						\Phi(\alpha):=\big(I+\alpha M_\s(q-1)\big)\big(I+\alpha M_\s(q-2)\big)\cdots\big(I+\alpha M_\s(0)\big)
					\end{equation}
					is semi-Schur and $\ker(\Phi(\alpha)-I)=\ker(M_\ave)$. Meanwhile, because the scheduling is periodic, $e((k+1)q)=\Phi(\alpha) e(kq)$ for all $k\in\N$. 
					
					Let $\tilde e=\begin{bmatrix}
						\tilde e_x\\\tilde e_\omega\\\tilde e_z
					\end{bmatrix}$ be the oblique projection of $e(0)=-\xi^*=-\begin{bmatrix}
						\hat x^*\\\hat\omega^*\\\frac{1}{\sqrt{k_I}}\hat L^{\frac{1}{2}}\lambda^*
					\end{bmatrix}$ onto $\ker(\Phi(\alpha)-I)=\ker(M_\ave)=\ker(M)$ along $\ran(\Phi(\alpha)-I)$. Corollary~\ref{cor:semi-Schur} in Section~\ref{sec:technical} implies that $e(kq)$ converges to $\tilde e$ exponentially fast as $k$ approaches to infinity. In fact, this also means that $e(k)$ converges to $\tilde e$ exponentially fast as $k$ approaches to infinity, since the error dynamics \eqref{discrete_time_LTV} is linear and periodic. 
					The rest of the proof is almost identical to the part in the proof of Theorem~\ref{thm:continuous_law}. It follows from $\tilde e\in\ker(M)$ and \eqref{KKT} that the pair $(\hat x^*+\tilde e_x, \hat\omega^*+\tilde e_\omega)$ is an optimal solution to the problem~\eqref{op_prob}, with an adjoint state $\hat\lambda\in\R^{pn}$ such that $\hat L^{\frac{1}{2}}\hat\lambda=\alpha k_I\sum_{k=0}^\infty\hat L_\s(k)^{\frac{1}{2}}\hat x(k)$ (such $\hat\lambda$ exists because of \eqref{ob_image_inclusion}). Finally, Lemma~\ref{lem:fundamental} and the exponential convergence $\lim_{k\to\infty}(\hat x(k),\hat\omega(k))=(\hat x^*+\tilde e_x, \hat\omega^*+\tilde e_\omega)$ imply that $\hat x(k)$ converges to a LS solution $x^*$ of~\eqref{LAE} in the sense of \eqref{hat_x^*&x^*} exponentially fast.
				\end{proof}
				
				{\color{blue}The proof of Theorem~\ref{thm:discrete_1} reveals that, for the DT-SD-LS algorithm to converge to a LS solution, the matrix $\Phi(\alpha)$ must be semi-Schur, and the condition $\ker(M_\ave) = \ker(M)$ must hold. A sufficient condition on the step size $\alpha$ that ensures semi-Schur $\Phi(\alpha)$ is provided by the formula~\eqref{def:alpha_max} in Lemma~\ref{lem:eigen_pert}. Regarding the latter condition, we note that periodic scheduling protocols beyond the cyclic one can also satisfy this requirement. Specifically, it follows directly from the definition of $M_{\ave}$ in~\eqref{def:bar_M} that if every portion of the solution is transmitted at least once within each period, then the equality $\ker(M_\ave) = \ker(M)$ is guaranteed.
				}
				
				Similar to Proposition~\ref{prop:x^*}, we can also characterize the LS solution given by the DT-SD-LS algorithm in the following proposition.
				
				\begin{proposition}\label{prop:x^*_2}
					There is a unique $\zeta^*\in\ran(\Phi(\alpha)-I)$ which is also an exact solution of the LAE~\eqref{KKT_matrx_0}. Moreover, The LS solution $x^*$ of \eqref{LAE} in Theorem~\ref{thm:discrete_1} found by Algorithm~\ref{alg:DT-SD-LS} is given by $\hat x^*=\begin{bmatrix}
						I_{np}&0_{(np)\times(2np)}
					\end{bmatrix}\zeta^*$ through \eqref{hat_x^*&x^*}.
				\end{proposition}
				
				Proposition~\ref{prop:x^*_2} can be proven similarly as Proposition~\ref{prop:x^*}, with the additional observation that $\ker(\Phi(\alpha)-I)=\ker(M_\ave)=\ker(M)$, and the difference that $\tilde e$ is now the oblique projection onto $\ker(M)$ along $\ran(\Phi(\alpha)-I)$, instead of onto $\ker(M)$ along $\ran(M)$.

				\section{Tracking LS solutions of LAEs with time-varying observation vectors}\label{sec:tracking}
				
				Contrast to the problem \eqref{LAE}, we consider the problem of solving 
				\begin{equation}\label{LAE_2}
					\underbrace{\begin{bmatrix}
							A_1\\\vdots\\A_p
					\end{bmatrix}}_{A}x=\underbrace{\begin{bmatrix}
							b_1(k)\\\vdots\\b_p(k)
					\end{bmatrix}}_{b(k)},
				\end{equation}
				where $b:\N\to\R^m$ is now a time-varying observation vector.
				Because the LS solution also becomes time-varying, solving~\eqref{LAE_2} for all $k\in\N$ is now a tracking problem. We would like to develop an algorithm which can track a LS solution of the time-varying LAE~\eqref{LAE_2}.

				Recall that when solving the static LAE problem~\eqref{LAE}, the observation vector $b$ is not explicitly used in the algorithm~\eqref{discrete-time-law_2-componentwise}; it is only used for the initialization of $y_{ij}$. Clearly, for time-varying observation vector, that approach becomes infeasible. We need to modify the update law for $y_{ij}$ in order to take the time-varying $b$ into account. Instead of \eqref{discrete-time-law_2-componentwise_y}, we let
				\begin{multline}\label{new_discrete_time_law_y}
					y_{ij}(k+1)=y_{ij}(k)+\hat A_{ij}(x_{ij}(k+1)- x_{ij}(k))\\-\alpha\sum_{\ell\in\Nc(j)}(y_{ij}(k)-y_{i\ell}(k))
					-(b_{ij}(k+1)-b_{ij}(k)).
				\end{multline}
				
				{\color{blue}The pseudo-code for the LS solution tracking problem is very similar to Algorithm~\ref{alg:DT-SD-LS} and is summarized here.}
				\begin{algorithm}[H]
					\caption{DT-SD-LS with time-varying observation vector}\label{alg:DT-SD-LS_2}
					{\color{blue}	\begin{algorithmic}[1]
							\Require $\alpha,k_I, k_P$.
							\State Initialize $x_{ij}(0)=0, y_{ij}(0)=-b_{ij}, z_{ij}(0)=0$ for all $i\in\cP, j\in\cQ$.
							\For{$k=0,1,\ldots,k_{\max}-1$, the $i$-th agent, $i\in\cP$}
							\For{$j=1,\ldots,q$}
							\If{$j=(k\bmod q)+1$}
							\State Broadcast $x_{ij}(k)$ to the $\ell$-th agent for all $\ell\in\cN(i)$.
							\State Update $x_{ij}(k+1), y_{ij}(k+1), z_{ij}(k+1)$ for all $j\in\cQ$ according to \eqref{discrete-time-law_1-componentwise_x}, \eqref{new_discrete_time_law_y}, \eqref{discrete-time-law_1-componentwise_z}.
							\Else
							\State Do not broadcast $x_{ij}(k)$.
							\State Update $x_{ij}(k+1), y_{ij}(k+1), z_{ij}(k+1)$ for all $j\in\cQ$ according to \eqref{discrete-time-law_1'-componentwise_x}, \eqref{new_discrete_time_law_y} and \eqref{discrete-time-law_1'-componentwise_z}.
							\EndIf
							\EndFor
							\EndFor
							\Ensure $x_i=\begin{bmatrix}
								x_{i1}\\\vdots\\x_{ip}
							\end{bmatrix}$, $i\in\cP$.
					\end{algorithmic}}
				\end{algorithm}
				
				The next theorem guarantees convergence of Algorithm~\ref{alg:DT-SD-LS_2}.
				
				\begin{theorem}\label{thm:track}
					Suppose that the matrix $\tilde M$ defined by \eqref{def:tilde_M} is diagonalizable for some $k_P,k_I$. For sufficiently small $\alpha>0$, there exist a time-varying LS solution to the LAE~\eqref{LAE_2} $x^*:\N\to\R^n$,  $x^*(k)=\begin{bmatrix}
						(x_1^*(k))^\top&\ldots&(x_q^*(k))^\top
					\end{bmatrix}^\top$, $x_j^*(k)\in\R^{n_j}$, and constants $\kappa_1,\kappa_2,\kappa_3>0$ such that each $x_{ij}(k)$ obtained by Algorithm~\ref{alg:DT-SD-LS_2} satisfy the inequality
					\begin{multline}\label{practical_stability}
						\Vert x_{ij}(k)-x_j^*(k)\Vert\leq \kappa_1e^{-\kappa_2k}\Vert \hat b(0)\Vert\\+\kappa_3\max_{l\in\N,l\leq k-1}\Vert \hat b(l+1)-\hat b(l)\Vert.
					\end{multline}
				\end{theorem}
				
				\begin{proof}
					Consider the discrete-time algorithm modified from \eqref{discrete_time_law_2}, by setting $\hat b$ time-varying:
					\begin{subequations}\label{discrete_time_law_3}
						\begin{align}
							\hat x(k+1)&=\hat x(k)-\alpha\Big(\hat A^\top(\hat A\hat x(k)-\hat b(k)-\Lc^{\frac{1}{2}}\hat\omega(k))\nonumber\\
							&\quad+k_P\hat L_\s(k)\hat x(k)+k_I\hat z(k)\Big),\\
							\hat \omega(k+1)&=\omega(k)+\alpha\Lc^{\frac{1}{2}}(\hat A\hat x(k)-\hat b(k)-\Lc^{\frac{1}{2}}\hat\omega(k)),\\
							\hat z(k+1)&=\hat z(k)+\alpha\hat L_\s(k)\hat x(k).
						\end{align}
					\end{subequations}
					Then the algorithm consisting of~\eqref{discrete-time-law_2-componentwise_x}, \eqref{new_discrete_time_law_y} and \eqref{discrete-time-law_2-componentwise_z} is exactly \eqref{discrete_time_law_3} written in agent-wise and portion-wise from, with the change of variables $\hat y(k):=\hat A\hat x(k)-\hat b(k)-\Lc^{\frac{1}{2}}\hat\omega(k)$.
					
					For each $k\in\N$, let $(\hat x^*(k), \hat \omega^*(k))\in\R^{pn}\times\R^{pn}$ be a solution to the problem~\eqref{op_prob} with $\hat b=\hat b(k)$. Then the KKT condition~\eqref{KKT} holds for some $\hat\lambda^*(k)\in\R^{pm}$. Equivalently,
					\begin{equation}\label{KKT_matrix}
						M\xi^*(k)=F\hat b(k),
					\end{equation}
					where $\xi^*(k):=\begin{bmatrix}
						\hat x^*(k)\\\hat\omega^*(k)\\\frac{1}{\sqrt{k_I}}\hat L^{\frac{1}{2}}\hat\lambda^*(k)
					\end{bmatrix}$, $M, F$ are defined in \eqref{algorithm_matrix_ct_ti}, \eqref{def:F}, respectively.
					Without loss of generality, we further let $\xi^*(k)$ be the \emph{minimal norm} solution of \eqref{KKT_matrix}; i.e.,
					\begin{equation}
						\xi^*(k)=M^\dagger F\hat b(k),
					\end{equation}
					where $M^\dagger$ is the pseudo-inverse of $M$.
					Define $c_0:=\Vert M^\dagger F\Vert$. We have
					\begin{align}
						\Vert\xi^*(k)\Vert&\leq c_0\Vert\hat b(k)\Vert,\label{bound_xi}\\
						\Vert\Delta\xi^*(k)\Vert&\leq c_0\Vert\Delta\hat b(k)\Vert,\label{bound_delta_xi}
					\end{align}
					where $\Delta\xi(k):=\xi(k+1)-\xi(k)$, $\Delta\hat b(k)=\hat b(k+1)-\hat b(k)$.
					
					Recall the matrix $\Phi(\alpha)$ defined in \eqref{def:Phi}. As discussed in the proof of Theorem~\ref{thm:discrete_1}, $\Phi(\alpha)$ is semi-Schur, so $\Phi(\alpha)-I$ is semi-Hurwitz. Let
					\begin{equation*}
						\Phi(\alpha)-I=\begin{bmatrix}
							V_1&V_0
						\end{bmatrix}\begin{bmatrix}
							\bar J\\&0
						\end{bmatrix}\begin{bmatrix}
							U_1\\U_0
						\end{bmatrix}=:VJU
					\end{equation*}
					be the Jordan decomposition of $\Phi(\alpha)-I$. 
								Meanwhile, let $\begin{bmatrix}
									E_{11}(k)&E_{12}(k)\\E_{21}(k)&E_{22}(k)
								\end{bmatrix}=E(k)=UM_\s(k)V$, where recall $M_\s(k)$ is defined in \eqref{algorithm_matrix_dt_tv} and $\ker(\Phi(\alpha)-I)=\ker(M_\ave)\subseteq\ker(M_\s(k))$, where $M_\ave$ is defined in \eqref{def:bar_M}. We have
								\begin{multline*}
									0=M_\s(k)V_1=VE(k)UV_1\\
									=V\begin{bmatrix}
										E_{11}(k)&E_{12}(k)\\E_{21}(k)&E_{22}(k)
									\end{bmatrix}\begin{bmatrix}
										0\\I
									\end{bmatrix}=V\begin{bmatrix}
										E_{12}(k)\\E_{22}(k)
									\end{bmatrix}.
								\end{multline*}
								Hence $E_{12}(k)=0, E_{22}(k)=0$ for all $k\in\N$. This implies 
								\begin{equation}\label{expand_M}
									M_\s(k)=V\begin{bmatrix}
										E_{11}(k)&0\\E_{12}(k)&0
									\end{bmatrix}U.
								\end{equation}

									
									Meanwhile, similar to \eqref{error_states}, define the error states
									\begin{align}
										e(k)=\begin{bmatrix}
											e_x(k)\\e_\omega(k)\\e_z(k)
										\end{bmatrix}:&=\begin{bmatrix}
											\hat x(k)-\hat x^*(k)\\
											\hat \omega(k)-\hat\omega^*(k)\\
											\sqrt{k_I}\tilde z(k)-\frac{1}{\sqrt{k_I}}\hat L^{\frac{1}{2}}\lambda^*(k)
										\end{bmatrix}\nonumber\\
										&=\begin{bmatrix}
											\hat x(k)\\
											\hat \omega(k)\\
											\sqrt{k_I}\tilde z(k)
										\end{bmatrix}-\xi^*(k),\label{error_state_new}
									\end{align}
									where $\tilde z$ is defined by \eqref{def:tilde_z}. 
									Since now $\hat x^*,\hat \omega^*, \hat \lambda^*$ are also time-varying, instead of \eqref{discrete_time_LTV}, the error dynamics becomes
									\begin{equation}\label{iterative_step}
										e(k+1)=(I+\alpha M_\s(k))e(k)-\Delta \xi^*(k).
									\end{equation}
									Let $a_1(k):=U_1e(k)$, $a_0(k):=U_0e(k)$ for all $k\in\N$. Multiply \eqref{iterative_step} by $U$ on the left, expand $M_s(k)$ by \eqref{expand_M},
									we conclude that
									\begin{subequations}
										\begin{align}
											a_1(k+1)&=\big(I+\alpha E_{11}(k)\big)a_1(k)-U_1\Delta \xi^*(k),\label{dt-sub-LTV}\\
											a_0(k+1)&=\alpha E_{12}(k)a_1(k)+a_0(k)-U_0\Delta\xi^*(k).
										\end{align}
									\end{subequations}
									Note that $V_1a_1$ is the oblique projection of $e$ onto $\ran(\Phi(\alpha)-I)$ along $\ker(\Phi(\alpha)-I)$. It has been shown in the proof of Theorem~\ref{thm:discrete_1} (and with the convergence guarantee in Lemma~\ref{lem:d_LTI})
									that when $b$ is time-invariant (i.e., $\Delta \xi^*(k)=0$ for all $k\in\N$), $e\to0$ exponentially fast, which in turn implies that $a_1\to 0$ exponentially fast. Hence when the input $\Delta \xi^*\equiv 0$, the discrete-time time-varying system given by \eqref{dt-sub-LTV} is globally exponentially stable~\cite[Definition 4.5]{HK:02}, which further implies by \cite[Lemma 4.6]{HK:02} that the system is also exponentially input-to-state stable. This means the existence of $c_1,c_2,c_3>0$ such that
									\begin{align*}
										\Vert a_1&(k)\Vert\leq c_1e^{-c_2 k}\Vert a_1(0)\Vert+c_3\Vert U_1\Vert \max_{j\in\N,j\leq k-1}\Vert\Delta \xi^*(k)\Vert\\
										&\leq c_1e^{-c_2 k}\Vert a_1(0)\Vert+c_0c_3\Vert U_1\Vert \max_{j\in\N,j\leq k-1}\Vert\Delta \hat b(k)\Vert\\
										&= c_1e^{-c_2 k}\Vert U_1\xi^*(0)\Vert+c_0c_3\Vert U_1\Vert \max_{j\in\N,j\leq k-1}\Vert\Delta \hat b(k)\Vert\\
										&\leq c_0c_1e^{-c_2 k}\Vert U_1\Vert\Vert \hat b(0)\Vert+c_0c_3\Vert U_1\Vert \max_{j\in\N,j\leq k-1}\Vert\Delta \hat b(k)\Vert
									\end{align*}
									where \eqref{bound_delta_xi}, $0$ initial state for $\hat x,\hat\omega,\tilde z$, and \eqref{bound_xi} are used for deriving the second, third, and forth lines, respectively. Recall the definition of $e$ in \eqref{error_state_new}. We conclude that
									\begin{multline}\label{time-varying_convergence}
										\left\Vert\begin{bmatrix}
											\hat x(k)\\
											\hat \omega(k)\\
											\sqrt{k_I}\tilde z(k)
										\end{bmatrix}-\xi^*(k)-V_0a_0(k)\right\Vert=\Vert V_1a_1(k)\Vert\\
										\leq c_0c_1e^{-c_2 k} \Vert V_1\Vert\Vert U_1\Vert\Vert\hat b(0)\Vert\\+c_0c_3\Vert V_1\Vert\Vert U_1\Vert \max_{j\in\N,j\leq k-1}\Vert\Delta \hat b(k)\Vert.
									\end{multline}
									Denote
									\begin{equation*}
										\xi^*(k)+V_0a_0(k)=:\begin{bmatrix}
											\tilde x^*(k)\\
											\tilde \omega^*(k)\\
											\tilde z^*(k)
										\end{bmatrix}.
									\end{equation*}
									Note that $V_0a_0(k)\in\ker(\Phi(\alpha)-I)=\ker( M_\ave)=\ker(M)$ for all $k\in\N$. Hence, $\xi^*(k)+V_0a_0(k)$ is also a solution to the KKT condition \eqref{KKT_matrix}. By Lemma~\ref{lem:fundamental}, \eqref{hat_x^*&x^*} holds and $x^*(k):=\begin{bmatrix}
										x_1^*(k)\\\vdots\\x_q^*(k)
									\end{bmatrix}$
									is a LS solution to the problem \eqref{LAE_2} for all $k\in\N$. We also conclude from \eqref{time-varying_convergence} that for any $i\in\cP,j\in\cQ$,
									\begin{multline*}
										\Vert x_{ij}(k)-x_j^*(k)\Vert\leq \left\Vert\begin{bmatrix}
											\hat x(k)\\
											\hat \omega(k)\\
											\sqrt{k_I}\tilde z(k)
										\end{bmatrix}-\xi^*(k)-V_0a_0(k)\right\Vert\\
										\leq c_0c_1e^{-c_2 k}\Vert V_1\Vert\Vert U_1\Vert\Vert \hat b(0)\Vert\\
										+c_0c_3\Vert V_1\Vert\Vert U_1\Vert \max_{l\in\N,l\leq k-1}\Vert\Delta \hat b(l)\Vert,
									\end{multline*}
									which gives \eqref{practical_stability} with $\kappa_1=c_0c_1\Vert V_1\Vert\Vert U_1\Vert$, $\kappa_2=c_2$ and $\kappa_3=c_0c_3\Vert V_1\Vert\Vert U_1\Vert$. This completes the proof.
								\end{proof}
								
								Note that unlike the time-invariant problem~\eqref{LAE}, when applying algorithm~\eqref{discrete_time_law_3} to solve \eqref{LAE_2}, $\xi^*(k)+V_0a_0(k)$ is no longer the the oblique projection of $\xi^*(k)$ onto $\ran(\Phi(\alpha)-I)$ along $\ker(\Phi(\alpha)-I)$. Therefore, while practical convergence to a time-varying solution trajectory $x^*(t)$ of \eqref{LAE_2} is assured, there is no explicit way to characterize $x^*(t)$ as done in Proposition~\ref{prop:x^*_2}.
									
									{\color{blue}\begin{remark}[On the tracking error]\label{rem:noisy_b}
											The input-to-state stability (ISS)-like estimate in~\eqref{practical_stability} for the tracking error is common in the dynamic consensus literature (see~\cite{SSK-BVS-JC-RAF-KML-SM:19}). While a persistent tracking error is unavoidable when the observation vector $b(k)$ changes monotonically with a constant rate, ISS ensures that if the rate of change of $b(k)$ converges, then the tracking error will also asymptotically vanish. On the other hand, if $b(k)$ is corrupted by noise, the single-step variation $\|\hat b(k+1) - \hat b(k)\|$ may be large, resulting in high uncertainty in the LS solution estimates. To address this, we suggest applying a low-pass filter to the signals $b_{ij}(k)$ before inputting them into Algorithm~\ref{alg:DT-SD-LS_2}. This smooths the variation in $b_{ij}(k)$ and thereby improves the accuracy of the LS solution estimates. Another approach is to use a time-varying sequence of damping factors to mitigate past errors, as explored in \cite{EM-JIM-CS-SM:14}. While this method reduces the tracking error for higher-order derivatives of the reference signal to some extent, it significantly increases algorithmic complexity, making it impractical for online implementation.
									\end{remark}}

									\section{Technical results}\label{sec:technical}
									
									In this section, we summarize several technical results that are utilized in the proof of the main theorems presented in this paper. To the best of the authors' knowledge, while some of these results are straightforward, they do not appear to be covered in the existing literature. Proving these results may also hold independent interest.
									
									\subsection{Semi-Hurwitz and semi-Schur matrices}\label{subsec:semi-Hurwitz}
									Recall the definitions of semi-Hurwitz and semi-Schur matrices as given in Definitions~\ref{def:semi-Hurwitz}, \ref{def:semi-Schur}.
									It is not difficult to see that a matrix $A\in\R^{n\times n}$ is semi-Hurwitz if it has a Jordan decomposition
									\begin{equation}\label{Jordan_decomp_c}
										A=\begin{bmatrix}
											V_1&V_0
										\end{bmatrix}\begin{bmatrix}
											J_1\\&0
										\end{bmatrix}\begin{bmatrix}
											U_1\\U_0
										\end{bmatrix}=VJU,
									\end{equation}
									where $J_1\in\R^{m\times m}$ for some $m\leq n$ is a block diagonal of Jordan blocks and is Hurwitz, $VU=I$. Similarly, a matrix $A\in\R^{n\times n}$ is semi-Schur if it has a Jordan decomposition
									\begin{equation}\label{Jordan_decomp_d}
										A=\begin{bmatrix}
											V_1&V_0
										\end{bmatrix}\begin{bmatrix}
											J_1\\&I
										\end{bmatrix}\begin{bmatrix}
											U_1\\U_0
										\end{bmatrix}=VJU,
									\end{equation}
									where $J_1\in\R^{m\times m}$ for some $m\leq n$ is a block diagonal of Jordan blocks and is Schur, $VU=I$. It also immediately follows from matrix perturbation theory that the matrix $I+\alpha A$ is semi-Schur if $A$ is semi-Hurwitz and $\alpha>0$ is sufficiently small.
									
									\begin{lemma}\label{lem:direct_sum}
										If $A\in\R^{n\times n}$ is semi-Hurwitz, then $\ran(A)\oplus\ker(A)=\R^n$. Similarly, if $A\in\R^{n\times n}$ is semi-Schur, then $\ran(A-I)\oplus\ker(A-I)=\R^n$.
									\end{lemma}
									\begin{proof}
										This is an immediate result of the Jordan decomposition \eqref{Jordan_decomp_c}, \eqref{Jordan_decomp_d}, such that $\ran(A)=\ran(V_1), \ker(A)=\ran(V_0)$ for semi-Hurwitz $A$, and $\ran(A-I)=\ran(V_1), \ker(A-I)=\ran(V_0)$ for semi-Schur $A$.
									\end{proof}
									
									Note that although we always have $\dim(\ran(A))+\dim(\ker(A))=n$, $\ran(A)\oplus\ker(A)=\R^n$ is not true in general. For example, for the $2\times 2$ matrix $A=\begin{bmatrix}
										0&1\\0&0
									\end{bmatrix}$, $\ran(A)=\ker(A)=\ran\left(\begin{bmatrix}
										1\\0
									\end{bmatrix}\right)$.
									
									Similar to the Lyapunov characterizations of Hurwitz matrices, we have the follow results for semi-Hurwitz matrices. Here, we use the notations $ A\succ 0, A\succeq 0, A\prec0, A\preceq 0$ to denote that the matrix $A$ is positive definite, positive semi-definite, negative definite, or negative semi-definite, respectively.
									
									\begin{lemma}\label{lem:Lyap_semi-Hurwitz}
										A matrix $A\in\R^{n\times n}$ is semi-Hurwitz if and only if there exist $P,Q\in\R^{n\times n}$, $P\succ 0, Q\succeq 0$ such that 
										\begin{equation}\label{Lyap_equation_c}
											PA+A^\top P+Q =0,
										\end{equation}
										and $A$ has no eigenvalues on the imaginary axis other than $0$.
									\end{lemma}
									
									\begin{proof}
										If $A$ is semi-Hurwitz, than clearly $A$ has no eigenvalues on the imaginary axis other than $0$. We now show the existence of $P\succ 0, Q\succeq 0$ satisfying the Lyapunov equation~\eqref{Lyap_equation_c}. Consider the Jordan decomposition~\eqref{Jordan_decomp_c}. Since the matrix $J_1$ is Hurwitz, there exist $P_1, Q_1\in\R^{m\times m}$, $P_1,Q_1\succ 0$ such that $P_1J_1+J_1^\top P_1+Q_1=0$. Let $P:=U^\top\diag(P_1,I_{n-m})U\succ 0$. We have
										\begin{multline*}
											PA+A^\top P=U^\top\left(\begin{bmatrix}
												P_1J_1\\&0
											\end{bmatrix}+\begin{bmatrix}
												J_1^\top P_1\\&0
											\end{bmatrix}\right)U\\
											=U^\top\begin{bmatrix}
												-Q_1\\&0
											\end{bmatrix}U=:-Q\preceq 0,
										\end{multline*}
										which proves \eqref{Lyap_equation_c}.
										
										To show the other direction, it first immediately follows from the properties of Lyapunov equation~\eqref{Lyap_equation_c} that all eigenvalues of $A$ must have non-positive real parts, and those eigenvalues on the imaginary axis must be non-defective. In addition, because $A$ has no eigenvalues on the imaginary axis other than $0$, $A$ is semi-Hurwitz and this completes the proof.
									\end{proof}
									The Lyapunov characterization and its proof for semi-Schur matrices is similar. We present it here without its proof.
									
									\begin{lemma}
										A matrix $A\in\R^{n\times n}$ is semi-Schur if and only if there exist $P,Q\in\R^{n\times n}$, $P\succ 0,Q\succeq 0$ such that
										\begin{equation}\label{Lyap_equation_d}
											A^\top P A-P+Q=0,
										\end{equation}
										and $A$ has no eigenvalues on the unit circle other than $1$.
									\end{lemma}
									
									Consider a continuous-time LTI system with input
									\begin{equation}\label{sys:c_LTI_u}
										\dot x(t)=Ax(t)+Bw(t),
									\end{equation}
									or a discrete-time LTI system with input
									\begin{equation}\label{sys:d_LTI_u}
										x(k+1)=Ax(k)+Bw(k).
									\end{equation}
									We show what $A$ being semi-Hurwitz or semi-Schur means to the stability of the systems \eqref{sys:c_LTI_u}, \eqref{sys:d_LTI_u}.
									
									\begin{lemma}\label{lem:c_LTI}
										Consider the continuous-time LTI system with input~\eqref{sys:c_LTI_u} and suppose that
										$A\in\R^{n\times n}$ is semi-Hurwitz, and $B\in\R^{n\times m}$ is such that $\ran(B)\subseteq\ran(A)$. Let $x^*$ be the oblique projection of $x_0\in\R^n$ onto $\ker (A)$ along $\ran(A)$, and let $x(t):\Rp\to\R^n$ be the solution of \eqref{sys:c_LTI_u} with initial state $x(0)=x_0$. Then, there exist $c_1,c_2,c_3>0$ such that
										\begin{equation}\label{c_ISS}
											\Vert x(t)-x^*\Vert\leq c_1e^{-c_2t}\Vert x_0-x^*\Vert + c_3\esssup_{s\in[0,t]}\Vert w(s)\Vert
										\end{equation}
										for all $t\in\Rp$.
									\end{lemma}
									Here, $\esssup$ mean the essential supremum.
									\begin{lemma}\label{lem:d_LTI}
										Consider the discrete-time LTI system with input~\eqref{sys:d_LTI_u} and suppose that $A\in\R^{n\times n}$ is semi-Schur, and $B\in\R^{n\times m}$ is such that $\ran(B)\subseteq\ran(A-I)$. Let $x^*$ be the oblique projection of $x_0\in\R^n$ onto $\ker (A-I)$ along $\ran(A-I)$, and let $x(k):\N\to\R^n$ be the solution of \eqref{sys:d_LTI_u} with initial state $x(0)=x_0$. Then, there exist $c_1,c_2,c_3>0$ such that
										\begin{equation}
											\Vert x(k)-x^*\Vert\leq c_1e^{-c_2k}\Vert x_0-x^*\Vert + c_3\max_{l\in\N,l\leq k}\Vert w(l)\Vert
										\end{equation}
										for all $k\in\N$.
									\end{lemma}
									
									We only prove Lemma~\ref{lem:c_LTI} here. The proof for Lemma~\ref{lem:d_LTI} is similar.
									
									\begin{proof}[Proof of Lemma~\ref{lem:c_LTI}]
										Let~\eqref{Jordan_decomp_c} be the Jordan decomposition of $A$. Since $\ran(B)\subseteq\ran(A)=\ran(V_1)$, $UB=\begin{bmatrix}
											U_1B\\0
										\end{bmatrix}$. Define $\begin{bmatrix}
											a_1(t)\\a_0(t)
										\end{bmatrix}:=Ux(t)$. Then $x(t)=V\begin{bmatrix}
											a_1(t)\\a_0(t)
										\end{bmatrix}$ and
										\begin{align*}
											\begin{bmatrix}
												\dot a_1(t)\\\dot a_0(t)
											\end{bmatrix}&=U\dot x(t)=UAV\begin{bmatrix}
												a_1(t)\\a_0(t)
											\end{bmatrix}+UBw(t)\\
											&=\begin{bmatrix}
												J_1a_1(t)+U_1Bw(t)\\0
											\end{bmatrix}.
										\end{align*}
										Hence $a_0$ is a constant vector, and $a_1$ is a solution to the linear differential equation $\dot a_1(t)=J_1a_1(t)+U_1Bw(t)$. Because $J_1$ is Hurwitz, this system is exponentially input-to-state stable, such that there exist $\kappa_1,\kappa_2,\kappa_3>0$ and $\Vert a_1(t)\Vert\leq \kappa_1 e^{-\kappa_2 t}\Vert a_1(0)\Vert+\kappa_3\Vert U_1\Vert\Vert B\Vert\esssup_{s\in[0,t]}\Vert w(s)\Vert$ for all $t\in\Rp$. Meanwhile, note that $x^*=V_0a_0$. Thus
										\begin{align*}
											\Vert x&(t)-x^*\Vert=\Vert V_1a_1(t)\Vert\leq \Vert V_1\Vert\Vert a_1(t)\Vert \\
											&\leq \Vert V_1\Vert \left(\kappa_1 e^{-\kappa_2 t}\Vert a_1(0)\Vert+\kappa_3\Vert U_1\Vert\Vert B\Vert\esssup_{s\in[0,t]}\Vert w(s)\Vert\right)\\
											&\leq \kappa_1\Vert V_1\Vert\Vert U_1\Vert  e^{-\kappa_2 t}\Vert V_1a_1(0)\Vert\\
											&\quad+\kappa_3\Vert V_1\Vert\Vert U_1\Vert\Vert B\Vert\esssup_{s\in[0,t]}\Vert w(s)\Vert\\
											&= \kappa_1\Vert V_1\Vert\Vert U_1\Vert  e^{-\kappa_2 t}\Vert x(0)-x^*\Vert\\
											&\quad+\kappa_3\Vert V_1\Vert\Vert U_1\Vert\Vert B\Vert\esssup_{s\in[0,t]}\Vert w(s)\Vert.\\
										\end{align*}
										We hence prove \eqref{c_ISS} with $c_1=\kappa_1\Vert V_1\Vert\Vert U_1\Vert$, $c_2=\kappa_2$, and $c_3=\kappa_3\Vert V_1\Vert\Vert U_1\Vert\Vert B\Vert$.
									\end{proof}
									
									Suppose that the LTI systems~\eqref{sys:c_LTI_u}, \eqref{sys:d_LTI_u} are autonomous; i.e., $B=0$. We immediately have the following results:
									
									\begin{corollary}\label{cor:semi-Hurwitz}
										Consider a continuous-time LTI system 
										\begin{equation}\label{sys:c_LTI}
											\dot x(t)=Ax(t)
										\end{equation}
										and suppose that
										$A\in\R^{n\times n}$ is semi-Hurwitz. Let $x^*$ be the oblique projection of $x_0\in\R^n$ onto $\ker (A)$ along $\ran(A)$, and let $x(t):\Rp\to\R^n$ be the solution of \eqref{sys:c_LTI} with initial state $x(0)=x_0$. Then, $x(t)$ converges to $x^*$ exponentially fast.
									\end{corollary}
									
									\begin{corollary}\label{cor:semi-Schur}
										Consider a discrete-time LTI system 
										\begin{equation}\label{sys:d_LTI}
											x(k+1)=Ax(k)
										\end{equation}
										and suppose that
										$A\in\R^{n\times n}$ is semi-Schur. Let $x^*$ be the oblique projection of $x_0\in\R^n$ onto $\ker (A-I)$ along $\ran(A-I)$, and let $x(t):\Rp\to\R^n$ be the solution of \eqref{sys:d_LTI} with initial state $x(0)=x_0$. Then, $x(k)$ converges to $x^*$ exponentially fast.
									\end{corollary}	
									
									\subsection{Proofs of the properties of the algorithm matrices}
									
									\begin{lemma}\label{lem:M}
										Consider a matrix $M\in\R^{(n+m)\times(n+m)}$, given by
										\begin{equation*}
											M=\begin{bmatrix}
												-\Theta\Theta^\top-a\Xi\Xi^\top&-b\Xi\\b\Xi^\top&0
											\end{bmatrix},
										\end{equation*}
										where $\Theta\in\R^{n\times l}$, $\Xi\in\R^{n\times m}$ and $a>0,b\in\R$. Then $M$ is semi-Hurwitz.
									\end{lemma}
									\begin{proof}
										Since
										\begin{align*}
											M+M^\top=-2\left(\begin{bmatrix}
												\Theta\Theta^\top&0\\0&0
											\end{bmatrix}+a\begin{bmatrix}
												\Xi\Xi^\top&0\\0&0
											\end{bmatrix}\right)\preceq 0,
										\end{align*}
										%
										the Lyapunov equation \eqref{Lyap_equation_c} is satisfied with $P=I_{n+m}, Q=-(M+M^\top)$. According to Lemma~\ref{lem:Lyap_semi-Hurwitz}, it is sufficient to show that $M$ has no eigenvalues on the imaginary axis other than $0$. To this end, let $\lambda\in\R$ be such that $\lambda j$ is an eigenvalue of $M$ associated with a non-zero eigenvector $v=\begin{bmatrix}
											v_1&v_2
										\end{bmatrix}$, $v_1\in\C^n,v_2\in\C^m$, where $j$ is the unit imaginary number. Let $v^*$ denote the complex conjugate of $v$. We have
										\begin{align*}
											\lambda j v^*v&=v^*M v\\
											&=-v_1^*\Theta\Theta^\top v_1-av_1^*\Xi\Xi^\top v_1-b(v_1^*\Xi v_2-v_2^*\Xi^\top v_1).
										\end{align*}
										Equivalently,
										\begin{equation}\label{real_vs_imag}
											-\Vert \Theta^\top v_1\Vert^2	-\Vert \Xi^\top v_1\Vert^2
											=\left(\lambda \Vert v\Vert^2+2b\imag(v_2^*\Xi^\top v_1)\right)j.
										\end{equation}
										Note that the left-hand side of \eqref{real_vs_imag} is real and non-positive, while its right-hand side is purely imaginary. Hence both sides must be $0$. Consequently, the left-hand side of \eqref{real_vs_imag} being $0$ implies $\Theta^\top v_1=0, \Xi^\top v_1=0$. With this observation, the right-hand side of \eqref{real_vs_imag} being $0$ implies $\lambda=0$, and this completes the proof.
									\end{proof}
									
									We present the following Bauer-Fike-like theorem for kernel-preserving perturbations.
									\begin{lemma}\label{lem:BF}
										Let $A,B\in\R^{n\times n}$ be matrices such that $A$ is diagonalizable and $\ker(A)\subset\ker(B)$. Then for any non-zero eigenvalue $\mu$ of $A+B$, it holds that
										\begin{equation}\label{BF}
											\min_{\lambda\in\Lambda(A)\backslash\{0\}}|\mu-\lambda|\leq\cond(V)\Vert B\Vert,
										\end{equation}
										where $\cond(\cdot)$ denotes the condition number of a matrix, and $V\in\R^{n\times n}$ is the transformation matrix for the diagonalization of $A$.
									\end{lemma}
									\begin{proof}
										Let
										\begin{equation*}
											A=VDV^{-1}=\begin{bmatrix}
												V_1&V_0
											\end{bmatrix}\begin{bmatrix}
												D_1\\&0
											\end{bmatrix}\begin{bmatrix}
												V_1&V_0
											\end{bmatrix}^{-1}
										\end{equation*}
										be the diagonalization of $A$ with $D_1\in\R^{m\times m}$ for some $m\leq n$, such that $\ker(A)=\ran(V_0)$. Thus for any $a\in\R^{n-m}$, $V_0a\in\ker(A)$. Meanwhile, let
										$\begin{bmatrix}
											E_{11}&E_{12}\\
											E_{21}&E_{22}
										\end{bmatrix}=E=V^{-1}BV$.
										Since $\ker(A)\subset\ker(B)$,
										\begin{equation*}
											0=B(V_0a)=VEV^{-1}V_0a=V\begin{bmatrix}
												E_{11}&E_{12}\\
												E_{21}&E_{22}
											\end{bmatrix}\begin{bmatrix}
												0\\a
											\end{bmatrix}=V\begin{bmatrix}
												E_{12}\\E_{22}
											\end{bmatrix}a
										\end{equation*}
										which in turn implies that $\begin{bmatrix}
											E_{12}\\E_{22}
										\end{bmatrix}=0$. As a result, the similarity transformation gives
										\begin{equation*}
											V^{-1}(A+B)V=\begin{bmatrix}
												E_{11}+D_1&0\\E_{21}&0
											\end{bmatrix}=:\tilde D.
										\end{equation*}
										Note that $\tilde D$ is a lower block triangular matrix. Hence $\Lambda(A+B)=\Lambda(\tilde D)=\Lambda(E_{11}+D_1)\cup\{0\}$.
										
										Now, for any $\mu\in \Lambda(E_{11}+D_1)$ and corresponding eigenvector $x\in\R^m$, the relation $(E_{11}+D_1)x=\mu x$ implies $E_{11}x= (\mu I_m-D_1)x$. Additionally, since $D_1$ is a diagonal matrix consists of elements in $\Lambda(A)\backslash\{0\}$, $\mu I_m-D_1$ is also a diagonal matrix and we further conclude that $\Vert E_{11}\Vert \Vert x\Vert\geq \Vert E_{11}x\Vert\geq \min_{\lambda\in\Lambda(A)\backslash\{0\}}|\mu-\lambda|\Vert x\Vert$. Therefore,
										\begin{multline*}
											\min_{\lambda\in\Lambda(A)\backslash\{0\}}|\mu-\lambda|\leq \Vert E_{11}\Vert\leq \Vert E\Vert=\Vert V^{-1}BV\Vert\\
											\leq \Vert V^{-1}\Vert \Vert B\Vert\Vert V\Vert, 
										\end{multline*}
										which proves \eqref{BF}.	 
									\end{proof}
									
									\begin{lemma}\label{lem:eigen_pert}
										Let $M_i\in\R^{n\times n}, i=1,\ldots,q$. Define $\bar M:=\frac{1}{q}\sum_{i=1}^q M_i$ and for any constant $\alpha>0$, define
										\begin{equation*}
											\Phi(\alpha):=(I+\alpha M_q)(I+\alpha M_{q-1})\cdots(I+\alpha M_1).
										\end{equation*}
										If $\bar M$ is diagonalizable and semi-Hurwitz, and $\ker(\bar M)\subseteq \ker(M_i)$ for all $i=1,\ldots,q$, then $\Phi(\alpha)$ is semi-Schur and $\ker(\Phi(\alpha)-I)=\ker(\bar M)$ for all {\color{blue}$\alpha\in(0,\max\{1,\alpha_{\max}\})$, where
											\begin{equation}\label{def:alpha_max}
												\alpha_{\max}:=-\hspace{-10pt}\max_{\lambda\in\Lambda(\bar M)\backslash\{0\}}\frac{2q\real(\lambda)}{\cond(V)\left(\sum_{\ell=2}^q\bar m^\ell\binom{q}{\ell}\right)\!+\!q^2|\lambda|^2}>0,
											\end{equation}	
											where $V\in\R^{n\times n}$ is the transformation matrix for the diagonalization of $\bar M$, and $\bar m:=\max_{i=1,\ldots,q}\Vert M_{i}\Vert$.
										}
									\end{lemma}
									\begin{proof}
										{\color{blue}Firstly, since $\bar M$ is semi-Hurwitz, $\real(\lambda)<0$ for all $\lambda\in\Lambda(\bar M)\backslash\{0\}$, which further proves that $\alpha_{\max}$ defined in \eqref{def:alpha_max} is positive.} Let $x\in\ker(\bar M)$. It follows from $\ker(\bar M)\subseteq \ker(M_i)$ that $(I+\alpha M_i)x=x$ for all $i=1,\ldots,q$. Thus $\Phi(\alpha)x=x$, which in turn implies $\ker(\bar M)\subseteq\ker(\Phi(\alpha)-I)$. Next, pick a non-zero vector $y\in\ker(\Phi(\alpha)-I)$ {\color{blue}and define $N(\alpha):=\Phi(\alpha)-I-\alpha q\bar M$. 
											By expanding the expression of $\Phi(\alpha)$, it can be obtained that
											\begin{equation}\label{expand_Phi}
												\Vert N(\alpha)\Vert \leq \sum_{\ell=2}^q(\alpha\bar m)^\ell\binom{q}{\ell}.
										\end{equation}}
										Meanwhile, because $\Phi(\alpha)y=y$, it follows from \eqref{expand_Phi} that 
										\begin{equation}\label{step_other_direction}
											\alpha q\bar My+N(\alpha)y=0.
										\end{equation}
										Suppose $y\not\in\ker(\bar M)$. Then $\Vert \bar My\Vert\geq \sigma\Vert y\Vert$, where $\sigma>0$ is the least non-zero singular value of $\bar M$. Pick $\alpha>0$ such that $\Vert N(\alpha)\Vert\leq \frac{\alpha q\sigma}{2}$. We have
										\begin{align*}
											\Vert \alpha q\bar My&+N(\alpha)y\Vert\geq \alpha q\Vert \bar My\Vert -\Vert N(\alpha)y\Vert\\
											&\geq \alpha q\Vert \bar My\Vert -\Vert N(\alpha)\Vert \Vert y\Vert\geq \frac{\alpha q\sigma }{2}\Vert y\Vert >0,
										\end{align*}
										which is a contradiction to \eqref{step_other_direction}. Hence we must have $y\in\ker(\bar M)$, which leads to $\ker(\Phi(\alpha)-I)\subseteq \ker(\bar M)$. As a result, we conclude $\ker(\Phi(\alpha)-I)=\ker(\bar M)$ when $\alpha$ is sufficiently small, which also indicates that the eigenvalue $1$ of $\Phi(\alpha)$ is non-defective because $\bar M$ is semi-Hurwitz. We now only need to show that the other eigenvalues of $\Phi(\alpha)$ are in the open unit disk, so that $\Phi(\alpha)$ is semi-Schur.
										
										To this end, let $\mu\in\Lambda(\Phi(\alpha))\backslash\{1\}$. Then it follows from the definition of $\Phi(\alpha)$ that $\mu-1\in\Lambda(\alpha q\bar M+N(\alpha))\backslash\{0\}$. It then follows from Lemma~\ref{lem:BF} that there exists $\lambda\in\Lambda(\bar M)\backslash\{0\}$ such that $|\mu-1-\alpha q\lambda|\leq \cond(V)\Vert N(\alpha)\Vert$. {\color{blue}By substituting the bound \eqref{expand_Phi} in,
											\begin{align*}
												|\mu|&\leq |\mu-1-\alpha q\lambda|+|1+\alpha q\lambda|\\
												&\leq \cond(V)\Vert N(\alpha)\Vert +\sqrt{(1+\alpha q\real(\lambda))^2+(\alpha q\imag(\lambda))^2}\\
												&\leq \cond(V)\left(\sum_{\ell=2}^q(\alpha\bar m)^\ell\binom{q}{\ell}\right)\\
												&\quad+\sqrt{1+2\alpha q\real(\lambda)+(\alpha q |\lambda|)^2}.
											\end{align*}
											Under the condition that $\alpha<\max\{1,\alpha_{\max}\}$, where $\alpha_{\max}$ is defined in \eqref{def:alpha_max}, we conclude that $|\mu|<1$. This completes the proof.
										}
										
										%
										%
									\end{proof}

									
									\section{Simulation examples}\label{sec:simulation}
									In this section, we validate our DT-SD-LS algorithm through two examples. The first example involves solving an overdetermined LAE, where we compare Algorithm~\ref{alg:DT-SD-LS} with several state-of-the-art distributed LS algorithms. In the second example, we apply Algorithm~\ref{alg:DT-SD-LS_2} to track a LS solution of an underdetermined LAE with a time-varying observation vector.
									
									{\color{blue}We compare the DT-SD-LS algorithm with the following distributed LS algorithms:
										\begin{enumerate}
											\item The implicit-explicit iteration from \cite{XW-JZ-SM-MJC:19},
											\item The gradient-tracking algorithm from \cite{TY-JG-JQ-XY-JW:20},
											\item The Arrow-Hurwicz-Uzawa (AHU) flow from \cite{YL-CL-BDOA-GS:19},
											\item The double-layered network approach from \cite{YH-ZM-JS:22}.
										\end{enumerate}	
										We test 6 overdetermined LAEs with $m = 256$ and $n = 32, 64, 96, 128, 160$, and $192$. Each problem admits a unique LS solution, denoted by $x^*$. The bandwidth is fixed at $\bar n = 16$. For all algorithms except the double-layered network, we use a circular topology with 8 agents. For the double-layered network, the first layer has 8 agents in a circular graph, while the second layer is fully connected.
										All methods except for the implicit-explicit iteration depend on a step-size parameter. Since no closed-form expressions for optimal step sizes exist, we empirically choose stable values for fair comparison. For the DT-SD-LS algorithm, we set $k_P = 25$ and $k_I = 5$.}
									
									{\color{blue}In terms of communication, the first two algorithms require transmitting both the guessed solution and an auxiliary vector of equal dimension at each iteration, so the number of communication cycles per computation grows linearly with the problem size under limited bandwidth. The double-layered network distributes this load by assigning components to different agents, fixing the communication cycles per iteration to 2. In contrast, the DT-SD-LS algorithm only transmits a part of the guessed solution, requiring just 1 communication cycle per iteration--offering the highest communication-to-computation (C-C) ratio. The setup details are summarized in Table~\ref{tab:1}.}
									
									{\color{blue}\begin{table}[h]
											\centering
											\begin{tabular}{|c||c|c|c|}
												\hline
												Algorithms &  Agents & Step size & C-C ratio \\ \hline\hline
												\textbf{Imp.-exp.} & 8 & - & $n/8$\\ \hline
												\textbf{G-tracking} & 8 & 0.001 & $n/8$  \\ \hline
												\textbf{AHU flow} & 8 & 0.001 & $n/8$\\ \hline
												\textbf{2-Layered} & $n/2$ & 0.01 & 2 \\\hline
												\textbf{SD-LS} & 8 & 0.01 & 1  \\ \hline
											\end{tabular}
											\caption{Comparison of different distributed LS algorithms.}	\label{tab:1}
									\end{table}}
									
									{\color{blue}
										Define 	
										\begin{equation}\label{def:e_1}
											e_1(k):=\frac{1}{p}\sum_{i=1}^p\Vert x_i(k)-x^*\Vert,
										\end{equation}
										which evaluates the convergence to the LS solution, and
										\begin{equation}\label{def:e_2}
											e_2(k):=\frac{1}{p}\sum_{i=1}^p\Vert x_i(k)-x_{\ave}(k)\Vert,
										\end{equation}
										where $x_\ave(k):=\frac{1}{p}\sum_{i=1}^p x_i(k)$. The quantity $e_2$ evaluates consensus among agents. The evolution of $e_1,e_2$ over communication cycles for all 6 LAEs is shown in Fig.~\ref{fig:compare}
									}
									
									\begin{figure}
										\centering
										\subfigure[$n=32$]{\input{figures/subplot_2_new.tex}\label{subfig:2}}
										\subfigure[$n=64$]{\input{figures/subplot_4_new.tex}\label{subfig:4}}
										
										\subfigure[$n=96$]{\input{figures/subplot_6_new.tex}\label{subfig:6}}
										\subfigure[$n=128$]{\input{figures/subplot_8_new.tex}\label{subfig:8}}
										
										\subfigure[$n=160$]{\input{figures/subplot_10_new.tex}\label{subfig:10}}
										\subfigure[$n=192$]{\input{figures/subplot_12_new.tex}\label{subfig:12}}
										\caption{Evolution of errors for the first example. Number of communication cycles is given by the horizontal axis. Evolution of $e_1$ is represented by solid curves, and evolution of $e_2$ is represented by dashed curves. Different colors represent curves from different algorithms: {\color{mycolor1} explicit-implicit iteration}, {\color{mycolor2} gradient tracking}, {\color{mycolor3}AHU flow}, {\color{mycolor4}double-layered network}, and {\color{mycolor5}SD-LS algorithm}.}\label{fig:compare}
									\end{figure}
									
									\begin{table}[h]
										\centering
										\begin{tabular}{|c||c|c|c|c|c|c|}
											\hline
											\multirow{2}{*}{Algorithms} & \multicolumn{6}{c|}{$n$} \\ \cline{2-7}
											& 32 & 64 & 96 & 128 & 160 & 192 \\ \hline\hline
											\textbf{Imp.-exp.} & 13226 & 30860 & 54500 & 76152 & $>10^5$ & $>10^5$\\ \hline
											\textbf{G-tracking} & \bf{2404} & \bf{8800} & \bf{21744} & 53708 & $>10^5$ & $>10^5$\\ \hline
											\textbf{AHU flow} & $>10^5$ & $>10^5$ & $>10^5$ & $>10^5$ & $>10^5$ & $>10^5$ \\ \hline
											\textbf{2-Layered} & $>10^5$ & 96992 & 76954 & 60620 & 61822& 88678\\ \hline
											\textbf{SD-LS} & 22645 & 26653 & 31663 & \bf{34068} & \bf{39879} & \bf{59519}\\ \hline
										\end{tabular}
										\caption{Number of communication cycles for $e_1$ to reach $10^{-4}$.}
										\label{tab:2}
									\end{table}
									
									{\color{blue}It is observed that even with a step size as small as $0.001$, the AHU flow does not converge well within $10^5$ communication cycles. For smaller problems ($n=32, 64$ and $96$), the gradient-tracking algorithm converges fastest in both LS error and consensus. However, as the problem size increases ($n= 128, 160$, and $192$), its performance degrades and is surpassed by the double-layered network and DT-SD-LS algorithm. For these larger problems, the proposed DT-SD-LS algorithm achieves the fastest convergence. Table~\ref{tab:2} shows the number of communication cycles needed for $e_1$ to reach $10^{-4}$. 
										This demonstrates DT-SD-LS's advantage in handling large problems with high communication efficiency. Although not yet optimized, further improvements in parameter tuning and rate analysis could significantly reduce iterations. We aim to explore this in future work, including applications to real-world large-scale LAEs.
									}

									\section{Conclusion}\label{sec:conclusion}
									
									In this paper, we proposed a novel DT-SD-LS algorithm for solving LAEs under limited bandwidth constraints. By introducing a periodic scheduling protocol, we ensure that only a portion of each agent's state vector is transmitted during each iteration, making the algorithm scalable and efficient even as the solution's dimension increases. Additionally, we addressed the scenario of time-varying observation vectors, demonstrating that our modified DT-SD-LS algorithm can track an LS solution trajectory, with the tracking error bounded by the single-step variation in the observation vector.
									
									{\color{blue}In future work, we aim to identify the optimal selection of the tunable parameters and step size that ensure the fastest convergence rate of the algorithm. Given the distributed nature of the problem, this identification may require decentralized or adaptive methods. We also plan to consider practical scenarios where communication is asynchronous or random, or where the network is time-varying due to potential communication failures or interruptions. Furthermore, we intend to test our algorithm on real-world large-scale problems.}

									\section*{Acknowledgment}
									The author would like to thank Professor Sonia Martinez at the University of California, San Diego, for the enjoyable and enlightening discussions that inspired this work. Special thanks also go to Professor Xuan Wang at George Mason University for suggesting the use of integrators in the distributed algorithm to facilitate constraint achievement.
									
									\bibliographystyle{IEEEtran}
									\bibliography{../bib/alias.bib,../bib/JC.bib,../bib/Main-add.bib,../bib/Main.bib}
									
									\vspace{-.7cm}
									\begin{IEEEbiography}[{\includegraphics[width=1in,height=1.25in,clip,keepaspectratio]{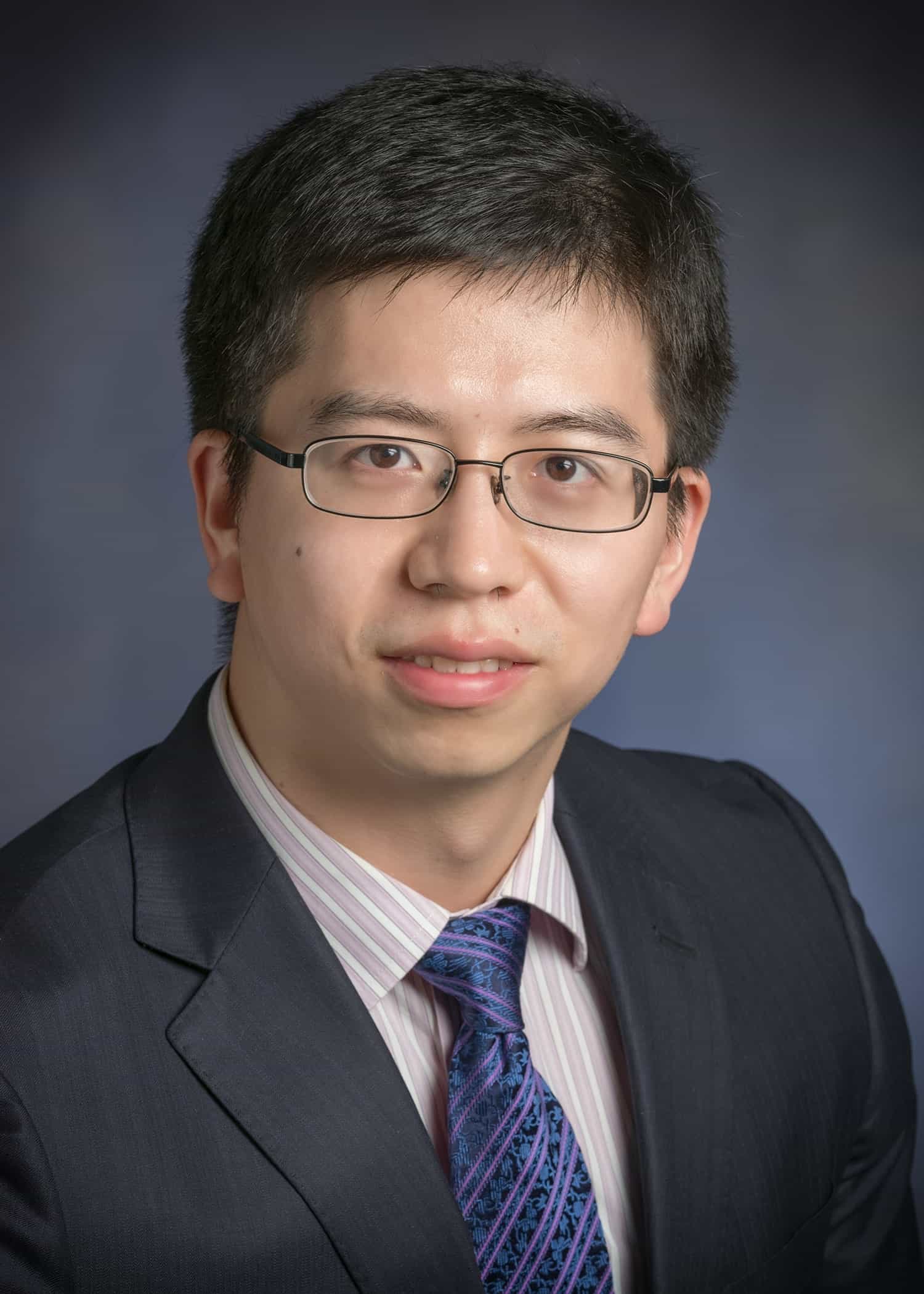}}]{Shenyu Liu} received his B. Eng. degree in Mechanical Engineering and B.S. degree in Mathematics from the National University of Singapore, Singapore, in 2014. He received his M.S. degree in Mechanical Engineering from the University of Illinois at Urbana-Champaign, in 2015, where he also received his Ph.D. degree in Electrical and Computer Engineering in 2020. He then spent two years in the Department of Mechanical and Aerospace Engineering at University of California, San Diego, as a postdoctoral researcher. Since 2022, he is an assistant professor in the School of Automation at Beijing Institute of Technology, Beijing, China. His current research interest includes stability theory of switched/hybrid systems, Lyapunov methods for nonlinear systems, matrix perturbation theory, distributed algorithms, and data-driven approach.
									\end{IEEEbiography}
									
								\end{document}